\documentclass{article}

\author{Art B. Owen\\Stanford University}
\title{A randomized Halton algorithm in R}
\date{May 2017}

\usepackage[round]{natbib}
\usepackage{amsmath,amsfonts,amssymb}
\usepackage{booktabs}
\usepackage{graphicx}
\usepackage{verbatim}
\usepackage{algorithm}
\usepackage{algorithmic}

\newcommand{\bsu}{\boldsymbol{u}}
\newcommand{\bsx}{\boldsymbol{x}}

\newcommand{\rd}{\,\mathrm{d}}
\newcommand{\dustd}{\mathbb{U}}

\newcommand{\tmod}{\ \mathsf{mod}\ }

\newcommand{\wh}{\widehat}

\newcommand{\mc}{\mathrm{MC}}

\newcommand{\hk}{\mathrm{HK}}

\newcommand{\mse}{\mathrm{MSE}}

\newcommand{\dnorm}{\mathcal{N}}

\newcommand{\vol}{{\mathrm{vol}}}
\newcommand{\var}{{\mathrm{Var}}}
\newcommand{\e}{\mathbb{E}}

\newcommand{\phz}{\phantom{0}}

\renewcommand{\le}{\leqslant}
\renewcommand{\ge}{\geqslant}
\renewcommand{\emptyset}{\varnothing}

\newcommand{\cx}{{\mathcal X}}

\newcommand{\proglang}[1]{#1}

\newcommand{\pkg}[1]{{\fontseries{b}\selectfont #1}}

\begin{document}
\maketitle
\begin{abstract}
Randomized quasi-Monte Carlo (RQMC) sampling can bring orders of magnitude 
reduction in variance compared to plain Monte Carlo (MC) sampling. 
The extent of the efficiency gain varies from problem to problem and 
can be hard to predict. 
This article presents an \proglang{R} function {\tt rhalton}
that produces scrambled versions of Halton sequences. 
On some problems it brings efficiency gains of several thousand fold. 
On other problems, the efficiency gain is minor. The code is designed 
to make it easy to determine whether a given integrand will benefit 
from RQMC sampling.  
An RQMC sample of $n$ points in $[0,1]^d$
can be extended later to a larger $n$ and/or $d$. 
\end{abstract}

\section{Introduction}

This paper is about a method for numerically approximating integrals of the
form $\mu = \int_{[0,1]^d}f(\bsx)\rd\bsx$.
There has been considerable recent progress in 
quasi-Monte Carlo (QMC) and randomized quasi-Monte Carlo (RQMC) solutions
to this problem. 
See \cite{dick:kuo:sloa:2013} for an introduction to the area.
(R)QMC methods can bring orders of magnitude improvements in accuracy
compared to Monte Carlo (MC).  This effect is most valuable in high dimensional
settings where classical alternatives to MC \citep[Chapter 5]{davrab} suffer from a curse
of dimensionality.
For other high dimensional problems, (R)QMC brings only minor improvements.
The best way to determine whether these methods improve on MC for a specific
problem is to run them both on a perhaps smaller version of that problem.  
This paper presents
a very simple to use RQMC code based on randomized Halton points.
The algorithm is designed to be extensible in both sample size and problem dimension.

The simplest approach to computing $\mu$ is to use a Monte Carlo estimate
$$
\hat\mu = \hat\mu_{\mc}\equiv \frac1n\sum_{i=1}^nf(\bsx_i)
$$
for independent $\bsx_i\sim\dustd([0,1]^d)$. This estimate converges almost
surely to $\mu$ by the law of large numbers, and when $\sigma^2=\int (f(\bsx)-\mu)^2\rd\bsx<\infty$,
then  $\sqrt{n}(\hat\mu-\mu)$ converges in distribution to $\dnorm(0,\sigma^2)$ 
by the central limit theorem.

The more general problem of finding the expectation of $g(\bsx)$ for $\bsx\sim p$
can often be solved by taking transformations $\tau(\bsx)\sim p$ for $\bsx\sim\dustd([0,1]^d)$.
\cite{devr:1986} contains a wealth of such transformations. Therefore we consider integration on the finite dimensional cube with the understanding that taking $f=g\circ \tau$ extends the range of application.

(R)QMC sampling can be far more effective than MC on some problems.
These problems tend to involve some smoothness on the part of the
integrand as well as a tendency for the integrand to be approximately
a superposition of lower dimensional integrands.  See~\cite{cafl:moro:owen:1997}
or \cite{sloa:wozn:1998}.
The presently known theory provides an explanation, in hindsight, for cases
where (R)QMC is seen to work well.  It is difficult to use the theory
prospectively.  Unlike MC where the root mean squared error is $\sigma n^{-1/2}$ for
all $n\ge1$, the convergence rates in QMC are based on asymptotic and worst case analyses
where non-trivially large powers of $\log(n)$ appear. We may see
good results for finite $n$ in some cases where the theory suggests otherwise
(e.g., functions with infinite variation).  In other settings, a given integrand $f$
can belong to multiple reproducing kernel Hilbert spaces, that each have
quite different consequences for integration accuracy.


This paper presents an RQMC method that for each random seed is conceptually
a doubly infinite random matrix $\cx$ with elements $\cx_{ij}$
for integers $1\le i<\infty$ and $1\le j<\infty$.
It is simple to implement.  We provide an implementation in R \citep{rlang:2015}.
There, the user typing {\tt rhalton(n,d)} gets a matrix $X\in[0,1]^{n\times d}$
comprising the upper left $n\times d$ submatrix of $\cx$, so
$X_{ij} = \cx_{ij}$ for $i=1,\dots,n$ and $j=1,\dots,d$.
The $i$'th row of $X$ is $\bsx_i = (x_{i1},\dots,x_{id})$
and then one gets $\hat\mu =(1/n)\sum_{i=1}^n f(\bsx_i)$.

To replicate the process one can call {\tt rhalton} repeatedly.
It is a good practice to set the seed prior to each call, so that
the computations are reproducible.
If $n$ observations are not enough, then it is possible
to skip the first $n$ observations and get the next ones
starting at the $n+1$'st.  To do that properly requires
taking some care with random seeds. Syntax to handle seeds
within {\tt rhalton} is deferred to Section~\ref{sec:design}.
There are also cases where one wants to extend a simulation
to a higher dimensional one. For instance, we may have simulated
a stochastic process $n$ times for $d$ time steps each using 
values from {\tt rhalton(n,d)}.  If we then want to continue the
simulation to $d'>d$ time steps,
we need colunns $d+1,\dots,d'$ of $\cx$. Extending to higher
dimensions without recomputing the first $d$ dimensions,
is also described in Section~\ref{sec:design}.

An outline of this paper is as follows.
Section~\ref{sec:qmc} introduces quasi-Monte Carlo concepts with
pointers to the literature for more information.
Section~\ref{sec:rqmc} presents randomized quasi-Monte Carlo which
can be used to get sample driven error estimates for QMC. In some cases,
randomization also improves accuracy and mitigates the large powers  of $\log(n)$ 
that appear in some performance metrics.
Section~\ref{sec:halton} presents the Halton sequences and various
randomizations and deterministic scrambles of them, including
a digital randomization that we adopt. The randomization that
we adopt was used in the simulations of \cite{wang:hick:2000}, although that paper
advocates a different randomization.
Section~\ref{sec:design} gives some design considerations for the 
implementation of {\tt rhalton} with examples of how to call it, using random
seeds to make simulations replicable as well as extensible in $n$ or $d$ or both.
Section~\ref{sec:illust} gives some numerical illustrations for problems with
dimension ranging from $1$ to $50$.  For some integrands we see an RMSE
matching theoretically predicted behavior close to $O(n^{-1})$ and a mild
dimension effect. Then RQMC is hundreds or even thousands of times as efficient
as plain Monte Carlo.
For other integrands there, the benefits of RQMC are smaller and decay rapidly with dimension.
Section~\ref{sec:comparison} makes a numerical  comparison
to the randomized Halton code function {\tt ghalton}
from the \proglang{R} package {\tt qrng} of \cite{hofe:lemi:2016}.
That code uses much more sophisticated scrambles. It is about $12$ to $24$ percent
more efficient on the example cases we consider.  The comparative advantages of {\tt rhalton}
are that it is extensible without recomputing intermediate
values, and that its pseudocode is so simple that it can be easily understood and
coded in new settings.

\section{Quasi-Monte Carlo}\label{sec:qmc}

In quasi-Monte Carlo sampling \citep{dick:pill:2010} we choose $n$ points $\bsx_1,\dots,\bsx_n\in[0,1]^d$ strategically
to make the discrete distribution of $\bsx_i$ for $i\sim\dustd\{1,2,\dots,n\}$ as close as possible
to the continuous $\dustd([0,1]^d)$ distribution. The difference between two such distributions is
known as a discrepancy and many discrepancies have been defined.  The most basic one is the
star discrepancy
\begin{align*}
D_n^*(\bsx_1,\dots,\bsx_n) & = \sup_{\bsx\in[0,1]^d}|\delta(\bsx)|,\quad\text{where}
\quad
\delta(\bsx)  = \frac1n\sum_{i=1}^n1_{\bsx_i\le\bsx} - \vol( [0,\bsx])
\end{align*}
is known as the `local discrepancy'.  This $\delta(\bsx)$ is the difference between the
fraction of points $\bsx_i$ in the box $[0,\bsx]\subset[0,1]^d$ and the fraction that box
should have gotten, which is simply its volume.
Then $D_n^*$ is defined in terms of the box with the greatest
mismatch between volume and fraction of points.  If $d=1$,  then $D_n^*$ is the
Kolmogorov-Smirnov distance between the empirical distribution of the $\bsx_i$ and the $\dustd([0,1])$ distribution.
Other discrepancy measures use different collections of sets,
or replace the supremum over sets by an $L_p$ measure or both.

If $D_n^*\to0$ and $f$ is Riemann integrable, then $\hat\mu\to\mu$
as $n$ increases. This is the QMC counterpart to the law of large numbers.
With genuine random numbers, the plain MC estimate works for
Lebesgue integrable functions, and so at first sight the QMC result
looks to be restrictive.  However pseudo-random numbers are
constructed in finite precision and so MC does not handle all Lebesgue integrable
functions either.

Next we turn to  the QMC counterpart to the central limit theorem.
For QMC we replace the concept of variance by variation. Let $V_{\hk}(f)$ be the total
variation of $f$ in the sense of Hardy and Krause.  For $d=1$, this is the
ordinary total variation from calculus.  
Multidimensional variation is more complicated and there have been many
generalizations. See \cite{variation} for a discussion and some historical notes.
The Koksma-Hlawka inequality is
\begin{align}\label{eq:kokhla}
|\hat\mu-\mu| \le D_n^*(\bsx_1,\dots,\bsx_n)\times V_{\hk}(f).
\end{align}
If we knew $D_n^*$ and $V_{\hk}$ then~\eqref{eq:kokhla} would provide a
100\% confidence interval for $\mu$ centered on $\hat\mu$.
In practice, $D_n^*$ can be expensive 
to compute and $V_{\hk}$ is almost certain to be harder to compute
than $\mu$ itself. We will address practical error estimation below.

Some low discrepancy constructions generate $\bsx_1,\dots,\bsx_n$
for a sequence of values $n$, such as $n=2^m$, along which
$D_n^*(\bsx_1,\dots,\bsx_n) = O( \log(n)^{d-1}/n)$.
The points in the $2^m$ point rule of the sequence are not necessarily among
the points of the $2^{m+1}$ point rule.  There are extensible
constructions of points $\bsx_1,\bsx_2,\dots$ where
$D_n^*(\bsx_1,\dots,\bsx_n) = O( (\log n)^dn^{-1})$ holds
along the entire sequence.  That is, one can attain extensibility
at the cost of an asymptotic logarithmic factor.

Given the rate at which discrepancy decreases, we find that
$|\hat\mu-\mu| =O( n^{-1+\epsilon})$ holds for any $\epsilon>0$,
when $V_{\hk}(f)<\infty$.  This rate establishes the potential
for QMC to be much more accurate than MC which has a 
root mean squared error of $O(n^{-1/2})$.
Even modestly large powers $d$ suffice to
make $\log(n)^d/n\gg n^{-1/2}$
for sample sizes $n$ of practical interest.
The logarithmic powers are usually not seen in applications.
For one thing, equation~\eqref{eq:kokhla} applies even to
a worst case function $f$ chosen based on the specific
points $\bsx_i$ in the integration rule.
The bound in~\eqref{eq:kokhla} is tight in that $V_{\hk}(f)$ cannot be
replaced by $(1-\eta)V_{\hk}(f)$ for any $\eta>0$, but it is
loose in that it can severely overestimate the error.
Another complication is that the implied constant in~\eqref{eq:kokhla}
has a strong dimensional dependence.
The first published bound grew very quickly with increased dimension $d$
and then \cite{atan:2004} proved a surprising result that
a still sharper bound rapidly decreases with $d$.
See \cite{faur:lemi:2009} for some comparisons.

The conservatism of~\eqref{eq:kokhla} can be partially understood
through a decomposition of $f$.  Let
$f(\bsx) = \sum_{u\subseteq1{:}d}f_u(\bsx)$ where
$f_u(\bsx)$ depends on $\bsx$ only through $x_j$ for $j\in u$.
The functions $f_u$ may come from an ANOVA decomposition
\citep{hoef:1948,sobo:1967:tran} or an anchored decomposition
\citep{rabi:etal:1999}.
Let $D_{n,u}^*$ be the $|u|$-dimensional star discrepancy of the points $\bsx_{i,u}$.  
Then
\begin{align}
|\hat\mu-\mu| &\le \sum_{u\subseteq1{:}d,u\ne\emptyset}\,
\Biggl| \frac1n\sum_{i=1}^nf_u(\bsx_i)-\int f_u(\bsx)\rd\bsx\Biggr|
\label{eq:decomp1}\\
&\le  \sum_{u\subseteq1{:}d, u\ne\emptyset} D_{n,u}^*\times V_{\hk}(f_u).
\label{eq:decomp2}
\end{align}
It is always true that $D^*_{n,u}\le D_n^*$.  The best QMC
constructions have very low discrepancies in their coordinate
projections, and typically $D^*_{n,u} = O( \log(n)^{|u|}/n)$.
If $f$ is dominated by low dimensional contributions,
then $V_{\hk}(f_u)$ for large $|u|$ could be negligible.
Then the error can then be like what one would see
in a low dimensional QMC applied to $f_u$.

Equation \eqref{eq:decomp2} is conservative.
\cite{moro:cafl:1994} have a similar expression
that is tighter than~\eqref{eq:decomp2}
because they use Vitali variation in place of Hardy-Krause.
Furthermore if $\sup_{\bsx}|f_u(\bsx)|<\epsilon$, then even if the Vitali variation
of $f_u$ is infinite, the error contribution of $f_u$ in~\eqref{eq:decomp1}
is below $2\epsilon$ which could be negligible
compared to $O((\log(n))^{|u|}/n)$ for the $n$ in use.

It is very hard to tell from looking at a functional form whether
the integrand $f$ is nearly a sum of functions of just a small
number of its input components.
\cite{cafl:moro:owen:1997}
find that a given $360$ dimensional function
designed to model a finance problem is very nearly a sum of
functions of one variable at a time.
It is possible to numerically inspect a function using Sobol' indices
to measure the extent to which it depends on just a few inputs.
It is even more simple to experiment on the function with
RQMC.

\section{Randomized quasi-Monte Carlo}\label{sec:rqmc}

One problem with QMC is that it is difficult to estimate
the error $|\hat\mu-\mu|$ because the points $\bsx_i$
are deterministic, and the bound~\eqref{eq:kokhla} is both
extremely conservative and much harder to compute than $\mu$ itself.

A practical remedy for this is to randomize the points $\bsx_i$.
In randomized quasi-Monte Carlo (see \cite{lecu:lemi:2002} for  a survey) 
we use points $\bsx_i\sim\dustd([0,1]^d)$ individually, that have
low discrepancy collectively.
Under randomized QMC,  $\e(\hat\mu)=\mu$ by uniformity
of the individual $\bsx_i$.
Also, if $V_{\hk}(f)<\infty$, then
$\var(\hat\mu) = O(n^{-2-\epsilon})$ for any $\epsilon>0$.
We can take a small number $R$ of independent randomizations
and pool them via
$$
\hat\mu = \frac1R\sum_{r=1}^R\hat\mu_r,\quad
\wh\var(\hat\mu) = \frac1{R(R-1)}\sum_{r=1}^R(\hat\mu_r-\hat\mu)^2.
$$
Then $\e(\hat\mu)=\mu$ and $\e(\wh\var(\hat\mu))=\var(\hat\mu)$,
so the variance estimate is not conservative.
The variance of the pooled estimate $\hat\mu$ is
$O(R^{-1}n^{-2-\epsilon})$ and it takes $nR$ function evaluations.
When accuracy of $\hat\mu$ is of primary importance
and the variance estimate is of secondary importance, 
then one takes a small value of $R$, perhaps only $5$ or $10$.
Larger values of $R$ are used in studies where the goal is to
measure how accurate RQMC is.

There are numerous randomization strategies.
The simplest is the Cranley-Patterson rotation
\citep{cran:patt:1976}.
Let $\lfloor z\rfloor$ be the greatest integer less than
or equal to $z$ and define $z\tmod 1=z-\lfloor z\rfloor$, the
fractional part of $z$.
Given a list of QMC points $\tilde\bsx_1,\dots,\tilde\bsx_n$,
Cranley and Patterson 
generate a (pseudorandom) vector $\bsu\sim\dustd([0,1]^d)$
and deliver $\bsx_i = \tilde\bsx_i  +\bsu \tmod 1$ (componentwise).
It is easy to see that $\bsx_i\sim\dustd([0,1]^d)$.  Also the points $\tilde\bsx_i$
that end up in a given axis-parallel rectangular subset of $[0,1]^d$
are those that started out as points $\bsx_i$ in the union of up to $2^d$
such subsets. This observation can be used to show that
$D_n^*(\bsx_1,\dots,\bsx_n)\le 4^d D_n^*(\tilde\bsx_1,\dots,\tilde\bsx_n)$.
We see at once that the convergence rates will be the same, but the
bound on the constant of proportionality is likely to be very conservative.
Fortunately we can use $\wh\var(\hat\mu)$ to estimate the actual error
magnitude and not a bound.

Cranley-Patterson rotations are well suited to QMC methods
known as lattice rules \citep{sloa:joe:1994}.
A second major category of QMC methods
use what are called digital constructions.
For sampling $[0,1]^d$, they include the Halton sequences
as well as the $(t,m,d)$-nets on $n=b^m$ points
and the extensible $(t,d)$-sequences of \cite{nied:1987}.
The remaining parameter $t\ge0$ is a quality parameter.
Smaller values indicate better equidistribution, though
the range of possible values for $t$ depends on $b$, $m$ and $d$.
This class of methods includes the sequences of
\cite{sobo:1967:tran} and \cite{faur:1982}.
Those algorithms generate observations
$\tilde x_{ij}=\sum_{\ell=1}^\infty \tilde a_{i,j}(\ell)b^{-\ell}$
for an integer base $b\ge2$ and digits $\tilde a\in\{0,1,\dots,b-1\}$.
The digits are usually chosen via the algebra of finite fields
to produce low discrepancy.
Digit scrambling methods then replace digits $\tilde a$
by  values $a=\pi(\tilde a)$ where $\pi$ is a random permutation
of $\{0,1,\dots,b-1\}$.
The permutation applied to $\tilde a_{i,j}(\ell)$ may depend on $j$
and $\ell$ and even on the digits $\tilde a_{i,j}(\ell')$ for $\ell'<\ell$.
Numerous such permutation strategies are described in~\cite{altscram}.

If there is a badly covered portion of $[0,1]^d$,
then Cranley-Patterson rotations simply shift the problem elsewhere.
Digit scrambling has the potential to improve upon QMC.
A digit scrambling from \cite{rtms} applied to $(t,d)$-nets
in base $b$ leads to a root mean squared error of $O( n^{-3/2}\log(n)^{(d-1)/2})$
along a sequence of values $n=\lambda b^m$
for $1\le\lambda<b$ and $m\ge0$, for smooth enough $f$.
It suffices for the mixed partial derivatives of $f$ taken at most once with respect to
each component $x_j$ to be in $L^2$. 
For any $f\in L^2[0,1]^d$, $\var(\hat\mu)=o(1/n)$, so that the
efficiency of RQMC versus MC increases to infinity with
no smoothness assumptions on $f$. 
Furthermore $\var(\hat\mu)   \le M\sigma^2/n$
where the constant $M$ depends on $t$, $d$ and $b$. 
For instance, if $t=0$ then $M\le \exp(1)$.
The logarithmic factors that might make the Koksma-Hlawka
bound much larger than $n^{-1/2}$ for feasible $n$, do not
make these RQMC root mean squared errors greatly exceed $n^{-1/2}$.

\section{Halton sequences}\label{sec:halton}

Halton sequences \citep{halt:1960}
presented here are easy to code and easy to understand.
Some QMC methods work best with specially chosen sample sizes such
as large prime numbers or powers of small prime numbers.
Halton sequences can be used with any desired sample size.
There may be no practical reason to require a richer
set of sample sizes than, for example, powers of $2$ \citep{sobo:1998}. However,
first time users may prefer powers of $10$, or simply the ability to 
select any sample size they like.

We begin with radical inverse sequences, following
the presentation in \cite{nied:1992}.
For integer $i\ge0$, write $i=\sum_{k=1}^\infty a_k(i)b^{k-1}$
for an integer base $b\ge2$ and digits $a_k(i)\in\{0,1,\dots,b-1\}$.
Because $i$ is finite, only finitely many $a_k(i)$ are nonzero.
Then the radical inverse function is
\begin{align}\label{eq:radinv}
\phi_b(i) = \sum_{k=1}^\infty a_k(i)b^{-k}
\end{align}
which is also a finite sum.

For $b=2$, the radical inverse sequence is due to \cite{vand:1935:I}.
Eight points of the
van der Corput construction are illustrated in Table~\ref{tab:vander}.
Integers $i$ are converted to base $2$ and then their digits are reflected
about the base $2$ `decimal' point, so for instance
$5\to101\to0.101\to0.625$ in base ten.
We see that these eight points generate values $\ell/8$ for $\ell=0,1,\dots,7$.
The sequence is commonly started at $0$, with $x_i=\phi_2(i-1)$
for $i=1,\dots,n$. Then if $n=2^m$ we will have
points $\ell/n$ for $\ell=0,1,\dots,n-1$, known as a left endpoint rule
\citep{davrab}.

\begin{table}\centering
\begin{tabular}{cccc}
\toprule
\multicolumn{2}{c}{$i$}
&\multicolumn{2}{c}{$x_i$}\\
\midrule
base 10 & base 2 & base 2 & base 10\\
\midrule
0 & \phz\phz0 & $.0$\phz\phz & $.0$\phz\phz\\
1 & \phz\phz1 & $.1$\phz\phz & $.5$\phz\phz\\
2 & \phz10 & $.01$\phz & $.25$\phz\\
3 & \phz11 & $.11$\phz & $.75$\phz\\
4 & 100 & $.001$ & $.125$\\
5 & 101 & $.101$ & $.625$\\
6 & 110 & $.011$ & $.375$\\
7 & 111 & $.111$ & $.875$\\
\bottomrule
\end{tabular}
\caption{\label{tab:vander}
$8$ consecutive points from the
radical inverse sequence in base $b=2$.
The base $2$ bits of $i$ are reflected about 
the (binary) decimal point to get those of $x_i$.
}
\end{table}

Because integers alternate between odd and even, the van der Corput
points alternate between $[0,1/2)$ and $[1/2,1)$. More generally,
for any $2^r$ consecutive integers $i$, the points $\phi_2(i)$
are stratified equally among intevals $[k2^{-r},(k+1)2^{-r})$ for 
$k\in\{0,1,\dots,2^r-1\}$, though not necessarily at the left endpoints.
Of $100$ consecutive points, the first $64$ will be equally distributed among
$64$ non-overlapping subintervals of length $1/64$, the next $32$ will be equally
distributed among $32$ intervals of length $1/32$ and the last $4$ will be
stratified over $4$ intervals of length $1/4$.
The star discrepancy of $n$ consecutive points in the van der Corput sequence is $O(\log(n)/n)$.  Along the subsequence
with $n=2^m$, for $m=0,1,2,\dots$, 
the star discrepancy is $O(1/n)$.

The same idea for low discrepancy sampling 
of $[0,1]$ works for other integer bases, where it is 
known as the generalized van der Corput construction.
The extension to $[0,1]^d$ was made by \cite{halt:1960}.
The Halton sequence has
$$
\bsx_i = (\phi_{p_1}(i-1),\phi_{p_2}(i-1),\dots,\phi_{p_d}(i-1))
$$
for different bases $p_j$, for $i=1,\dots,n$.
Here $p_j\ge2$ must be relatively prime integers.
Ordinarily $p_j$ is simply the $j$'th prime number, and this
is why we write $p_j$ instead of $b_j$.
The discrepancy of the Halton sequence is $O(\log(n)^d/n)$.

\begin{figure}
\centering 
\includegraphics[width=.9\hsize]{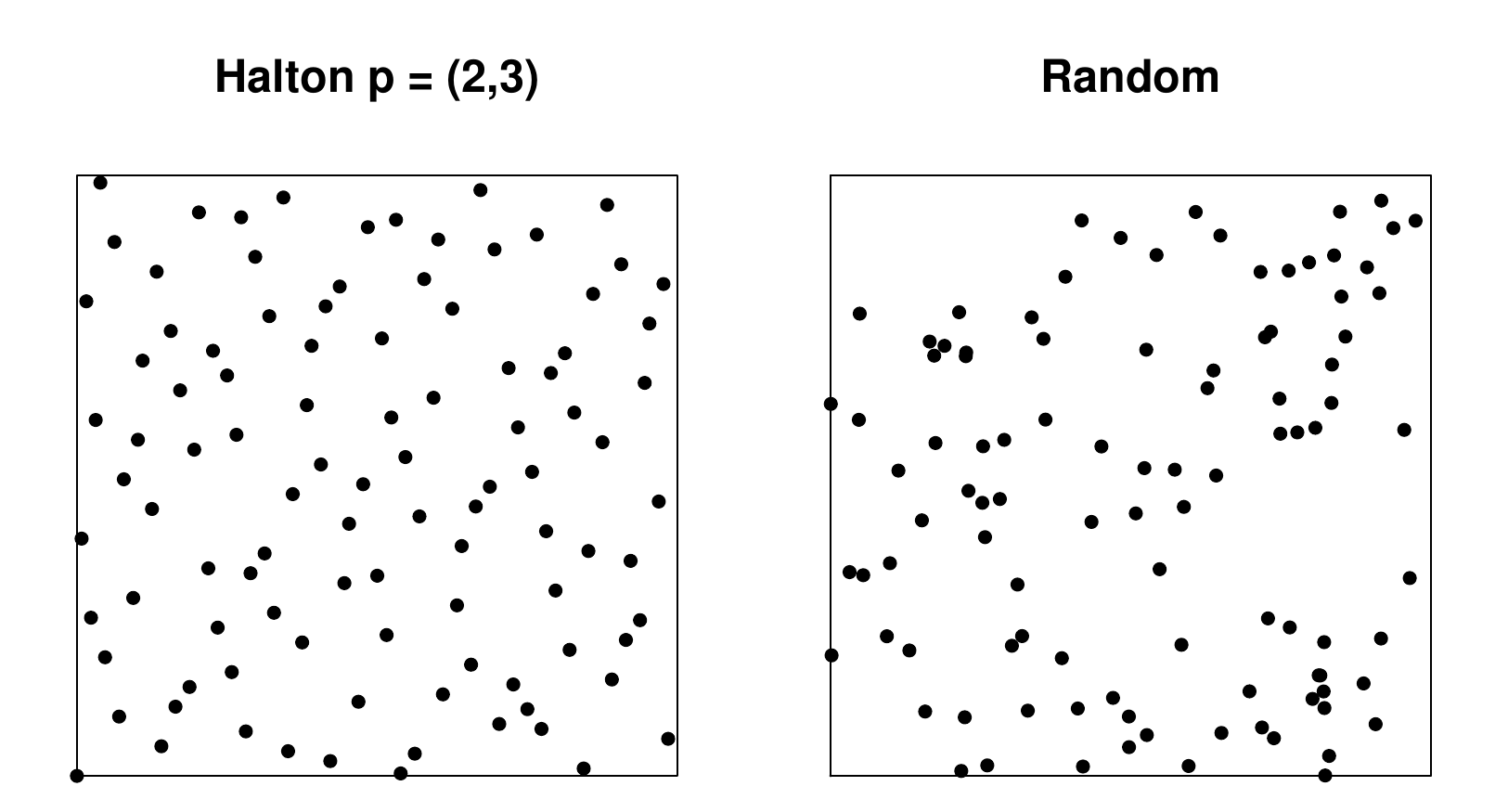}
\caption{\label{fig:haltonvsrandom}
The left panel shows the first $100$ Halton points in $[0,1]^2$. 
The right panel shows $100$ pseudorandom points. 
}
\end{figure}

Figure~\ref{fig:haltonvsrandom} compares the first $100$ points of the
Halton sequence in $[0,1]^2$ to $100$ pseudorandom points.
The Halton points show less tendency to form clumps and leave voids
than random points do.  

Notice that the first Halton point is at
$(0,0)$ which can be problematic in some uses, such as transformations
of $\dustd[0,1]$ random variables to $\dnorm(0,1)$ random variables via $\Phi^{-1}$.
The problem with $\bsx_1=0$ can easily be solved by starting the Halton
sequence at some larger index, such as $x_{ij} = \phi_{b_j}(N+i-1)$
for some large $N$.  Also, in a randomized QMC setting with
$\bsx_i\sim\dustd([0,1]^d)$ individually, points near the origin are not
a much more severe problem under RQMC than they would be with MC.

For large values of $d$ the Halton sequence has a problem.
To take an extreme case, suppose that $n<p_j$.
Then the first $n$ points in the sequence are $x_{ij} = (i-1)/p_j$.
The first $n$ points of $(x_{ij},x_{i,j+1})$ then lie along a diagonal
line with slope $p_{j}/p_{j+1}\doteq 1$ in the unit square. If $n$ is
a small multiple of $p_{j+1}$ then the points lie within a small number of such
parallel lines.
A Cranley-Patterson rotation would simply move the line or lines to a random
location in the square.

\begin{figure}
\centering 
\includegraphics[width=.9\hsize]{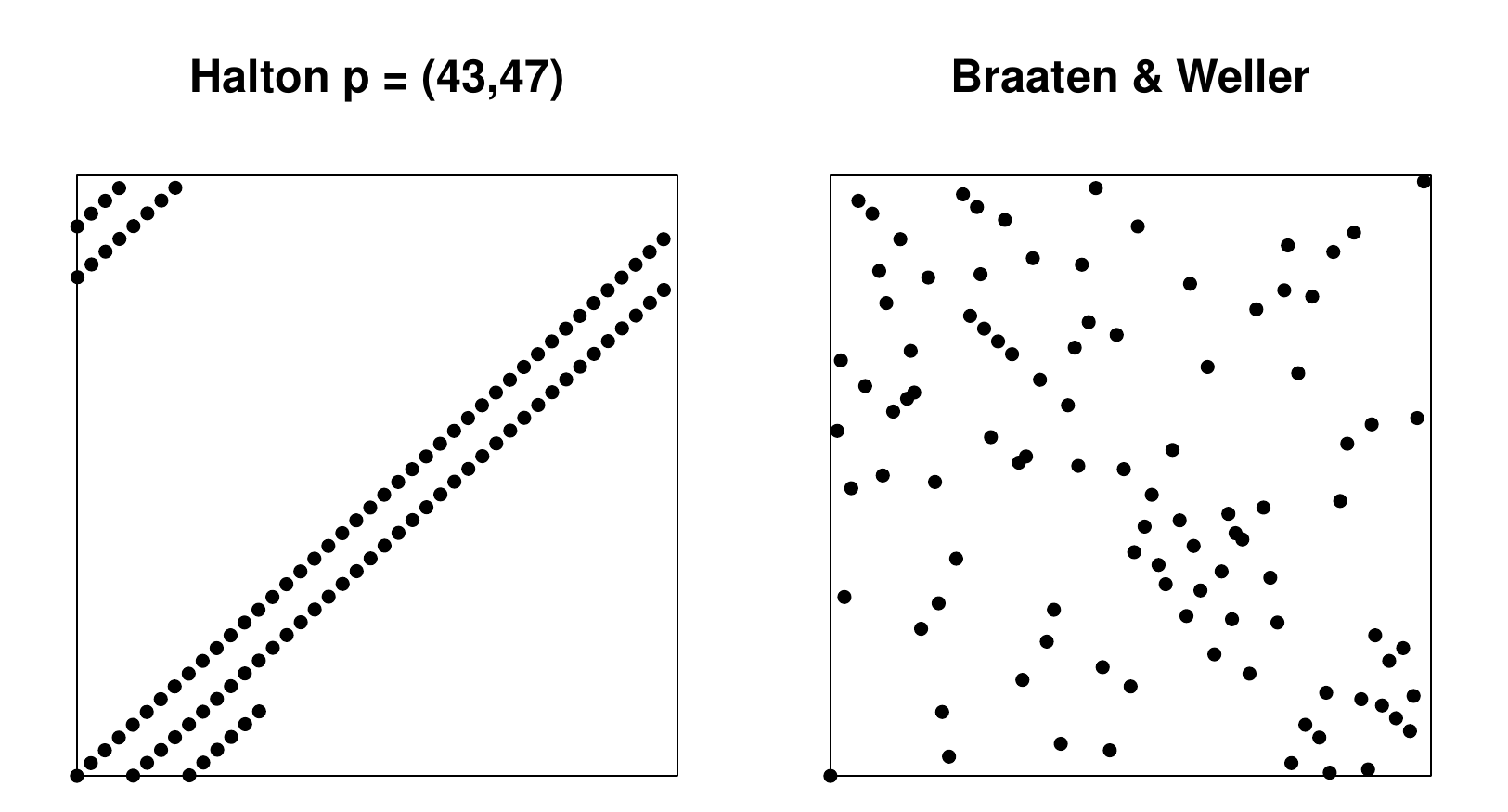}
\caption{\label{fig:deterscramhalton}
The left panel shows the first $100$ Halton points in $[0,1]^2$
for $p_{14}=43$ and $p_{15}=47$. 
The right panel shows $100$ pseudorandom points. 
}
\end{figure}

\cite{braa:well:1979} proposed a remedy to this striping problem.
For $i = \sum_{k=1}^\infty a_k(i,j)p_j^{k-1}$,
they replace
$$
x_{ij} = \sum_{k=1}^\infty a_k(i,j)p_j^{-k}
\quad\text{by}\quad
\sum_{k=1}^\infty \pi_j(a_k(i,j))p_j^{-k}
$$
for carefully chosen permutations $\pi_j$ of
$\{0,1,\dots,p_j-1\}$.
Their permutations have
$\pi_j(0)=0$.  Because $0$ is a fixed point in the permutation,
only finitely many digits are needed when computing each $x_{ij}$.
For each base $p_j$ of interest, they  chose $\pi_j(k)$
in a greedy step by step fashion to minimize the discrepancy of their
first $k$ choices.
Figure~\ref{fig:deterscramhalton} shows the first $100$ Halton points
in the $14$'th and $15$'th dimensions, along with the result of
Braaten and Weller's scrambling, using permutations from their Table 1.
We see that scrambling has broken up the stripe pattern in the Halton points instead of 
shifting it to another location. There does appear to remain an artifact with 
too many points near the diagonal $x_{i,14}+x_{i,15}=1$. 

\cite{faur:1992} presented an algorithm for choosing permutations
$\pi$ that have relatively good discrepancy properties compared
to other permutations especially the identity permutation.
Further permutations have been selected
by \cite{tuff:1998:b} as well as 
\cite{vand:cool:2006} who develop permutations with small
mean squared discrepancy and give a thorough survey
of permutation choices.
\cite{chi:masc:warn:2005} conduct a search for random
linear permutations of the form $\pi(a) = a\times b_j\tmod p_j$
for strategically chosen $b_j\in\{1,2,\dots,p_j-1\}$.
\cite{faur:lemi:2009} provide Matousek-style linear
scrambles of Halton sequences. Their scrambles
are deterministic and have $0$ as a fixed point.

Deterministic scrambles will not suit our present
purpose because we want to support replication based error assessment.
\cite{okte:shah:gonc:2012} investigate numerous deterministic
scrambling methods and exhibit conspicuous sampling artifacts in several
of them. They then advocate choosing the permutation
$\pi_j$ uniformly at random from the $p_j!$ permutations.
It appears that they include cases with $\pi_j(0)\ne0$.
Their Figure 3 shows that even for $p_{173}=1031$ and $p_{174}=1033$
there appear to be no serious sampling artifacts for $n=500$.

\begin{figure}
\centering 
\includegraphics[width=.9\hsize]{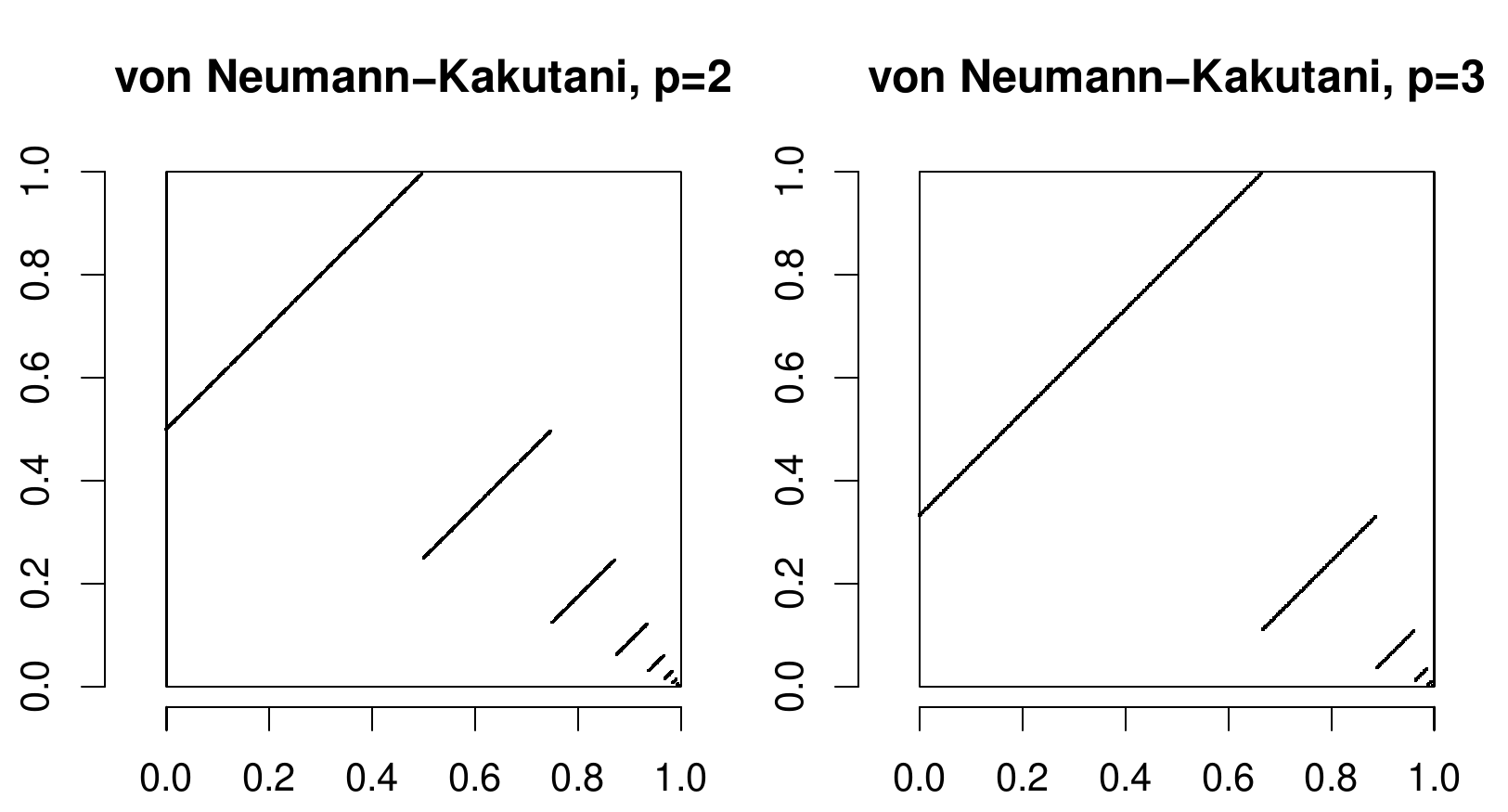}
\caption{\label{fig:vonnkaku}
The left panel shows $\phi_2(i+1)$ as a function of 
$\phi_2(i)$ for $i\ge0$. The right panel shows $\phi_3(i+1)$
versus $\phi_3(i)$. 
These are von Neumann-Kakutani transformations 
of $[0,1]$ to $[0,1]$. 
}
\end{figure}

\cite{wang:hick:2000} present an ingenious randomization
strategy for Halton sequences.  They choose a point
$\bsu\sim\dustd([0,1]^d)$.
Then writing $u_j \doteq \sum_{k=1}^Ka_kp_j^{-k}$ there is
an index $N=N_j = \sum_{k=1}^ka_kp_j^{k-1}$, they deliver
a stream of $x_{ij}= \phi_{p_j}(N_j+i-1)$ for $i\ge1$.  They do not have to actually
construct $N_j$ since the generalized van der Corput points
can be computed recursively uing the von Neumann-Kakutani
transformation depicted in Figure~\ref{fig:vonnkaku}.
Incidentally, it is interesting that the generalized van der Corput
sequences in the Halton sequence can all be started at different
$N_j$ if so desired.
Unfortunately, the random start Halton approach also
produces stripes as Figure 1 of 
\cite{chi:masc:warn:2005} shows for $n=512$ points
using primes $p_{13}=41$ and $p_{14}=43$.

A strategy based on a single random permutation $\pi_j$ for all the
bits in $x_{ij}$ will not suit our purposes. For instance,
with $p_1=2$, there are only two different permutations
$(0,1)$ and $(1,0)$.  In a $d$ dimensional problem there are
$\prod_{j=1}^dp_j!$ different permutations, which might seem
to be enough, except that having only two different permutations
for the first component will be a severe limitation when we employ $R>2$ replicates.
For any $i$, the only two possible values for $x_{i1}$ sum to 
binary $0.11111\dots=1$.
As a result, this strategy would not deliver $\bsx_i\sim\dustd([0,1]^d)$
upon which RQMC is based.

The points we use are formed by
\begin{align}\label{eq:permhalton}
x_{ij} = \sum_{k=1}^\infty \pi_{j,k}(a_k(i,j))p_j^{-k}
\end{align}
where $\pi_{j,k}$ are independent random permutations, each uniformly
distributed over all $p_j!$ possibilities.
This is one of the methods that
\cite{wang:hick:2000} included in their numerical comparisons.
It usually had greater efficiency in their higher dimensional
simulations (dimensions $10$, $20$ and $50$ from
Tables 2, 3 and 4) than the random start method had,
though random start did better for dimension $5$ (their Table 1).

It is clear from equation~\eqref{eq:permhalton}
that individually $x_{ij}\sim\dustd([0,1])$.
Also, if $j\ne j'$ then $x_{ij}$ and $x_{i'j'}$ have no permutations in common
and hence are independent, whether or not $i=i'$.
It then follows that $\bsx_i\sim\dustd([0,1]^d)$.
Equation~\eqref{eq:permhalton} involves an infinite number
of digits for each $x_{ij}$ because $\pi_{j,k}(0)$ is not
necessarily $0$. In floating point arithmetic one can
stop adding digits to $x_{ij}$ when $1.0-p_j^k<1.0$ no longer
evaluates to true.

\begin{figure}
\centering 
\includegraphics[width=.9\hsize]{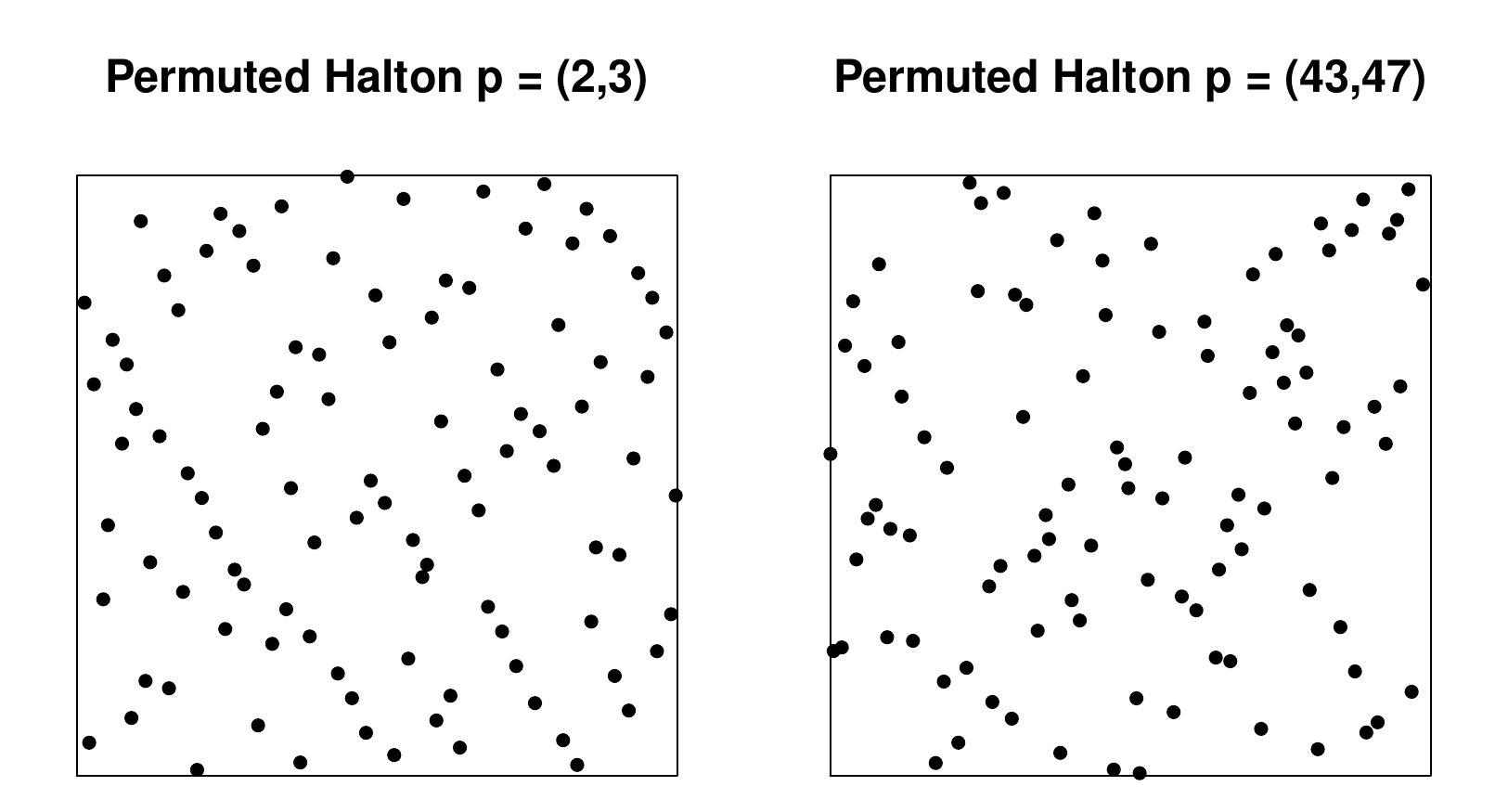}
\caption{\label{fig:randomscramhalton}
The left panel shows the first $100$ Halton points in $[0,1]^2$
for $p_1=2$ and $p_2=3$ after permutation via~\eqref{eq:permhalton}.
The right panel has the same points for $p_{14}=43$ and $p_{15}=47$. 
}
\end{figure}


\section{Design considerations}\label{sec:design}

The scrambling algorithm is illustrated in the pseudo-code 
of Algorithm~\ref{alg:rh}. The {\tt rhalton} function iterates 
over indices of the columns, invoking Algorithm~\ref{alg:rh}
on a set of indices given the $j$'th prime $p_j$ for
the $j$'th dimension.  It also keeps track of random seeds
and lets the user extend the output to more rows and/or more columns.
The greatest amount of \proglang{R}
code in the Appendix is the function {\tt nthprime} which 
returns the $n$'th prime from a list of $1000$ primes 
stored in the function.

If one wants to extend the function to an indefinitely large
number of primes, then one use the sieve of Eratosthenes
or Atkin's sieve on integers up to some maximum $D$.
The {\tt numbers} \proglang{R} package  \citep{borc:2017}
has a function {\tt atkin\_sieve} to do this.
It remains to choose $D$.
From~\cite{ross:1941} we know that $d$'th prime
is no larger than $D=d\log(d)+d\log(\log(d))$
if $6\le d\le \exp(95)\doteq1.81\times10^{41}$. This would remove the
need to store a list of primes and Rosser's result extends well 
beyond the practically relevant range of dimensions.  Our
implementation caps $d_0+d$ at $1000$ because
diminishing gains are expected for larger dimensions
and using a list means there are no package dependencies for {\tt rhalton}.

\begin{algorithm}[t]
\begin{algorithmic}
\STATE {b2r $\gets$ $1$ / $b$}
\STATE {ans $\gets$ ind $\times$ 0}
\STATE {res $\gets$ ind}
\WHILE  {1 $-$ b2r $<$ 1}
\STATE dig $\gets$ res $\bf{mod}$ b 
\STATE $\pi \gets$  uniform random permutation of $\{0,1,\dots,b-1\}$
\STATE pdig $\gets$ $\pi(\text{dig})$
\STATE ans $\gets$ ans + pdig $\times$ b2r 
\STATE b2r $\gets$ b2r / b 
\STATE res $\gets$ (res $-$ dig) / b 
\ENDWHILE 
\RETURN{ans}
\end{algorithmic}
\caption{Randomized van der Corput}\label{alg:rh}
\vspace*{.1cm}
NB: Input is a vector ind of non-negative integer indices and a prime base $b\ge2$. 
The while loop terminates in floating point arithmetic. If numbers are represented
some other way (e.g., rational numbers),
terminate when b2r $<\varepsilon$ for a small~$\varepsilon$ such as $10^{-16}$.
\end{algorithm}

Next we give a  summary of the design considerations in this code.
First, the baseline QMC methods was chosen to be digital 
instead of an integration lattice. Integration lattices 
require a search for suitable parameter values. While
that provides a way to tune them to a specific problem it
also places a burden on the user to know how to tune them.

It is likely that  the most accurate digital point sets are
$(t,m,d)$-nets, or for smoother integrands, higher order
digital nets \cite{dick:2011}.  Those constructions are
best used with sample sizes of the form $n=p^m$ for
prime numbers $p$. The Halton sequence by contrast
is less sensitive to special sample sizes. Also, it can
be applied in any dimension $d$.

The best variance for scrambled digital nets comes from the
nested uniform scramble introduced in \cite{rtms}
or from a partial derandomization of them due to \cite{mato:1998:2}.
Matousek's random linear scrambles are used in the \proglang{R} package
{\tt qrng} of \cite{hofe:lemi:2016}.
For those scrambles, the permutation applied to digit
$a_{i,j}(\ell)$ depends on $a_{i,j'}(\ell)$ for all $j'<j$. 
Random linear scrambles require much less additional
storage and bookkeeping than nested uniform ones do.
Even with that derandomization it remains
complicated to extend a prior simulation to increased $n$ or $d$.
The scramble in~\eqref{eq:permhalton} is simpler to use.  

The appendix has code for an \proglang{R}
function {\tt rhalton}. The most basic usage is
{\tt rhalton(n,d)} which yields an $n\times d$
matrix $X$ with $i$'th row $\bsx_i\in[0,1]^d$ of a Halton sequence.
If $f$ is a function on $[0,1]^d$ then 
{\tt mean(  apply(rhalton(n,d), 1,f))} provides an RQMC estimate
of $\mu = \int_{[0,1]^d}f(\bsx)\rd\bsx$.
Figure~\ref{fig:repexamp} shows an \proglang{R} code snippet
that provides replicated RQMC points.
It returns an estimate and standard error of
\begin{verbatim}
[1] 1.016707e+02 7.419982e-03
\end{verbatim}
where the true integral is $20^2/4+20/12\doteq 101.6667$,
about $0.5$ standard errors below the estimate.

\begin{figure}
\begin{verbatim}
f <- function(v){sum(v)^2} # Example function that is easy for RMQC 
R <- 10 
n <- 5000 
p <- 20 
stride <- 1000 # Or replace 1000 by nthprime(0,getlength=TRUE) 
muvec <- rep(0,R) 
for( r in 1:R ){
  x <- rhalton(n,p,singleseed = r*stride)  
  muvec[r] <- mean( apply( x,1,f ) ) 
}
mu <- mean(muvec) 
se <- sqrt(var(muvec)/R) 
print(c(mu,se)) 
\end{verbatim}
\caption{\label{fig:repexamp}
Example of replication with {\tt rhalton}. 
It estimates $\mu = \int_{[0,1]^p}f(\bsx)\rd\bsx$ for $p=20$
using $R=10$ replicates of $n=5000$ points. 
The initial seeds are separated by a `stride' that is larger than
the integrand's dimension to prevent overlap.
}
\end{figure}

Setting the seed as in the snippet of Figure~\ref{fig:repexamp}
makes it easier to do reproducible research.
A second goal for the user might be to extend a computation
later to larger values of $n$ and/or $d$.
To each integer value $s\ge0$ of the argument {\tt singleseed}
the scrambled Halton sequence  is a
matrix $\cx^{(s)}$ of values $\cx^{(s)}_{ij}$ for $1\le i<\infty$
and $1\le j<\infty$.
Calling {\tt rhalton(n,d,n0,d0,singleseed=s)}
will return an $n\times d$ matrix $X$ with
elements $X_{ij}=\cx^{(s)}_{n_0+i,d_0+j}$ for indices $i = 1,\dots,n$
and $j=1,\dots,d$.
Notice that $n_0$ is the number of rows you have already
and the first delivered new point will be the $n_0+1$'st one of $\cx^{(s)}$.
Similarly $d_0$ is the number of columns you have already
and the first delivered new variable will be the $d_0+1$'st one.
The {\tt rhalton} code has a fixed list of prime numbers.
It is necessary to have $d_0+d$ at most equal to the size
of that list (presently $1000$). If more columns are needed,
then the user can increase the length of the  list of prime numbers.

\begin{figure}[t!]
\centering
\begin{verbatim}
> rhalton(3,4,singleseed=1)                  # First example matrix
          [,1]       [,2]     [,3]      [,4]
[1,] 0.3581924 0.02400209 0.142928 0.7024411
[2,] 0.8581924 0.35733543 0.742928 0.1310125
[3,] 0.1081924 0.69066876 0.342928 0.2738697
> rhalton(5,4,singleseed=1)                  # Two more rows of the example
          [,1]       [,2]     [,3]      [,4]
[1,] 0.3581924 0.02400209 0.142928 0.7024411
[2,] 0.8581924 0.35733543 0.742928 0.1310125
[3,] 0.1081924 0.69066876 0.342928 0.2738697
[4,] 0.6081924 0.13511320 0.942928 0.9881554
[5,] 0.4831924 0.46844654 0.542928 0.4167268
> rhalton(2,4,n0=3,singleseed=1)             # Just the two new rows
          [,1]      [,2]     [,3]      [,4]
[1,] 0.6081924 0.1351132 0.942928 0.9881554
[2,] 0.4831924 0.4684465 0.542928 0.4167268
> rhalton(3,6,singleseed=1)                  # Two more columns
          [,1]       [,2]     [,3]      [,4]      [,5]      [,6]
[1,] 0.3581924 0.02400209 0.142928 0.7024411 0.2247080 0.5583664
[2,] 0.8581924 0.35733543 0.742928 0.1310125 0.5883443 0.8660587
[3,] 0.1081924 0.69066876 0.342928 0.2738697 0.7701625 0.1737510
> rhalton(3,2,d0=4,singleseed=1)             # Just the new columns
          [,1]      [,2]
[1,] 0.2247080 0.5583664
[2,] 0.5883443 0.8660587
[3,] 0.7701625 0.1737510
\end{verbatim}
\caption{\label{fig:seeding}
Extending {\tt rhalton} to larger $n$ and/or $d$, as described in the text.
}
\end{figure}

Figure~\ref{fig:seeding} shows how to extend
the output of {\tt rhalton}.
The first call to {\tt rhalton} generates $n=3$ points in
$d=4$ dimensions.  The next call generates $n=5$
points in $d=4$ dimensions.  Because the seed vector is the
same,  the first three rows of the second result 
contain the first result.  To just get the two new rows after
the third, the next example sets $n_0=3$.
The next call shows how to extend the matrix to get two
new columns including the first ones and the final call
shows how to get just the two new columns by setting
$d=2$ and $d_0=4$.

Columns 1 and 3 of the output matrices in Figure~\ref{fig:seeding} have a striking pattern
of common digits in their base $10$ representation.
This arises because the first and third prime
numbers are divisors of $10$.
The first two values in column one have the same second digit,
the first four have the same third digit, the first eight
have the same fourth digit and so on.
That pattern would not arise in a nested uniform scramble.

The default seeding uses the random seed $s+j-1$ for column
$j$ of $\cx^{(s)}$. That is, the user supplied seed is for column $1$
and all others are keyed off of that.
If the user does not want to employ consecutive random seeds for different
dimensions then it is possible to supply instead 
a parameter called {\tt seedvector} containing seeds $s_j$ for $j=1,\dots,d_0+d$.
It is not enough for the user to supply only $d$ seeds when $d_0>0$.
This design choice is meant to prevent an accidental misusage
in which some single seed $s_j$ is used for both dimension $j$ and $d_0+j$.

\section{Illustration}\label{sec:illust}

Here we illustrate the accuracy of scrambled Halton sequences
through some numerical examples.
Sometimes additive functions are used to test numerical
integration, but such functions are far too simplistic
for QMC. More commonly product functions are used.
Product functions are prone to have a coefficient of variation
that grows exponentially with dimension.
For this section, we choose some integrands over $[0,1]^d$ that resemble
quantities one sees in statistical applications and for which
$d$ can naturally vary.

These integrands are of the form
$$
f_k(\bsx) = g_k\Bigl(\frac1{\sqrt{d}}\sum_{j=1}^d\Phi^{-1}(x_j)\Bigr)
$$
for functions $g_k$ given below.
The argument to $g_k$ has a $\dnorm(0,1)$ distribution which makes it
easier to study properties of the integration problem.
The specific functions we choose are
$$
g_1(z) = \Phi(z+1),\quad g_2(z) = 1_{z+1\ge0},
\quad g_3(z) = \max(z+1,0) \quad\text{and}\quad
g_4(z) =  1_{z<\Phi^{-1}(0.001)}.
$$

Integrand $f_1$ is designed to make it easy for QMC to improve
on MC.  It is very smooth. The $+1$ inside $\Phi$ is
there so that this test function will not be antisymmetric.  Some QMC algorithms 
incorporate antithetic sampling and they would be exact for $g(z)=\Phi(z)$
by symmetry instead of by equidistribution.
Next, $V_{\hk}(f_2) = \infty$ for $d\ge2$. Yet in an ANOVA decomposition
of $f_2$ the lower order terms will dominate
\citep{grie:kuo:sloa:2010},
and they will be better behaved. For instance, only the full $d$ dimensional
interaction will be discontinous, the others will be smoother by integration.
Next, $V_{\hk}(f_3)=\infty$ for $d\ge3$, though it should also enjoy
the good projecton property from \cite{grie:kuo:sloa:2010}.
We can reasonably expect it to be more difficult than
$f_1$ but easier to handle than $f_2$.
Integrand $f_4$ describes a rare event, and it is included to show QMC failing to bring much improvement.
QMC is not designed for rare events; importance sampling is required, though it can be
combined with (R)QMC \citep{aist:dick:2014:tr}.

Each integral was estimated using randomized Halton points for a
range of dimensions and sample sizes $n$.
The true mean and variance of $f_k$ are
$\mu_k = \e(g_k(z))$ and $\sigma_k^2=\var(g_k(z))$, respectively, 
for $z\sim\dnorm(0,1)$, and these can be very accurately found
by applying $\Phi^{-1}$ to a midpoint rule on $10^7$ points in $[0,1]$.
The points in replicate $r$ were $\bsx_i^{(r)}\in[0,1]^d$.
We estimate the MSE for integrand $k$ by
$$
\wh\mse_k=\frac1R\sum_{r=1}^R(\hat\mu_{k,r}-\mu_k)^2,\quad
\hat\mu_{k,r} = \frac1n\sum_{i=1}^nf_k(\bsx_i^{(r)})
$$
We also retain the sample variance of the 
squared errors $(\hat\mu_{k,r}-\hat\mu_k)^2$ for $r=1,\dots,R$.
Those sample variances  enable an estimate of $\var(\wh \mse_k)$.
We suppress the dependence on $n$  and $d$
of $\mse_k$ and $\wh\mse_k$.

The efficiency of RQMC to MC is estimated by
$(\sigma^2_k/n)/\wh\mse$.
The uncertainty in $\mse$ is judged by lower and upper limits
$$
\frac{\sigma^2_k/n}{\wh\mse + 2\sqrt{\wh\var(\wh\mse)/R}},
\quad\text{and}\quad
\frac{\sigma^2_k/n}{\wh\mse - 2\sqrt{\wh\var(\wh\mse)/R}}.$$

\begin{figure}[t!]
\centering 
\includegraphics[width=.9\hsize]{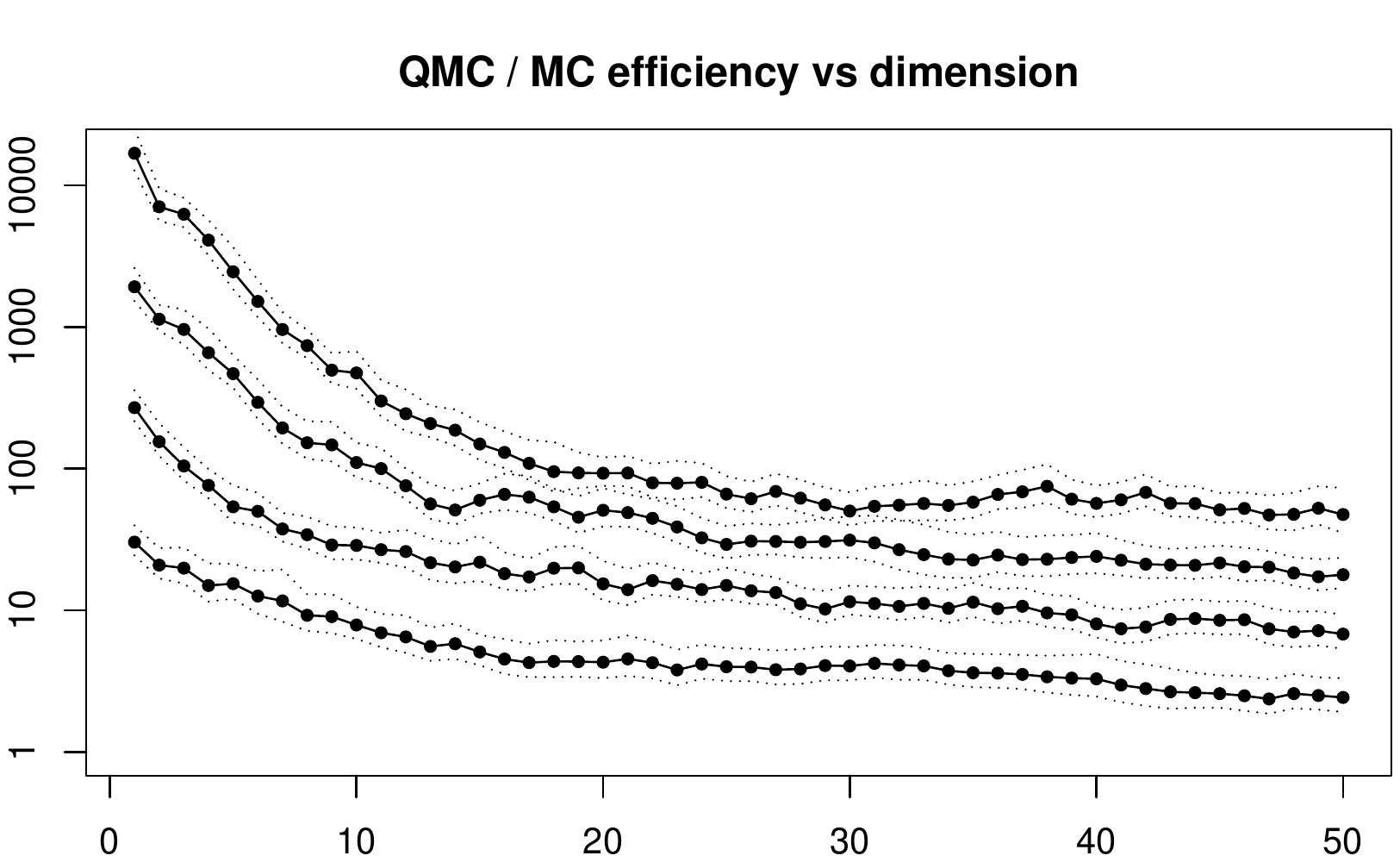}
\caption{\label{fig:case1}
RQMC efficiency for function $f_1$ based on $R=100$ repetitions.
Top to bottom $n=10^5$, $10^4$, $10^3$, $10^2$.
}
\end{figure}

\begin{figure}
\centering 
\includegraphics[width=.9\hsize]{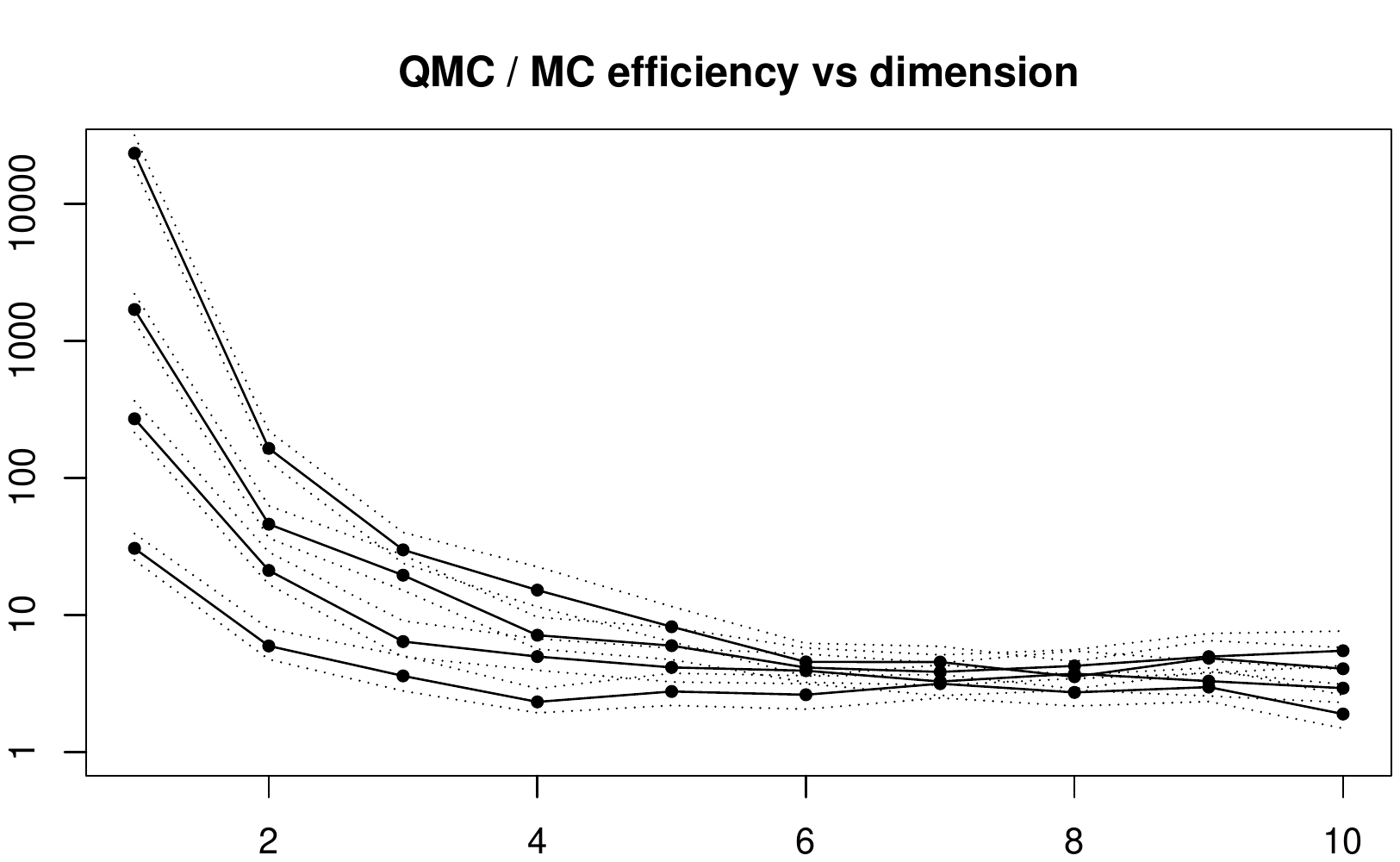}
\caption{\label{fig:case2}
RQMC efficiency for function $f_2$ based on $R=100$ repetitions.
Top to bottom $n=10^5$, $10^4$, $10^3$, $10^2$.
}
\end{figure}

\begin{figure}
\centering 
\includegraphics[width=.9\hsize]{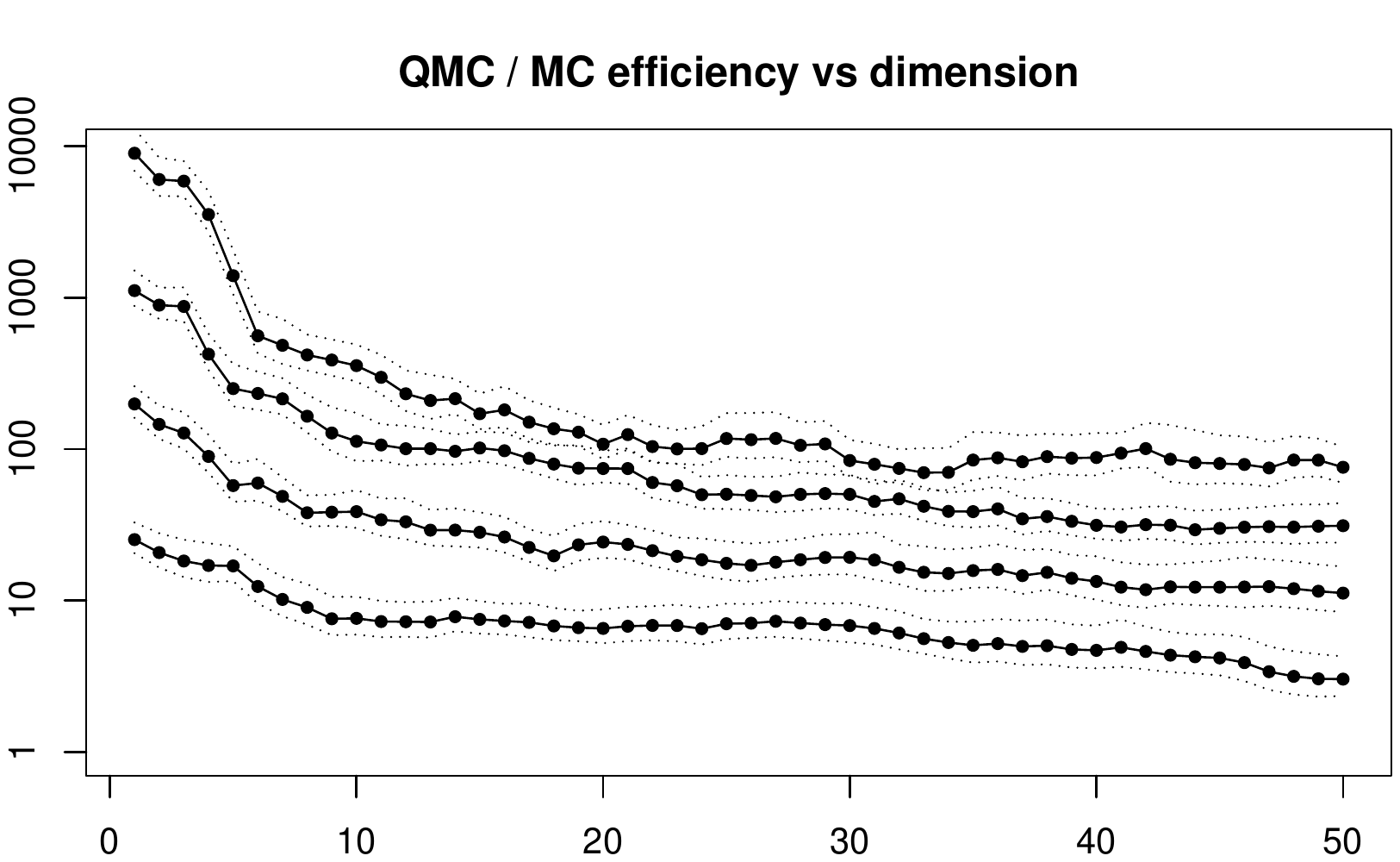}
\caption{\label{fig:case3}
RQMC efficiency for function $f_3$ based on $R=100$ repetitions.
Top to bottom $n=10^5$, $10^4$, $10^3$, $10^2$.
}
\end{figure}

\begin{figure}
\centering 
\includegraphics[width=.9\hsize]{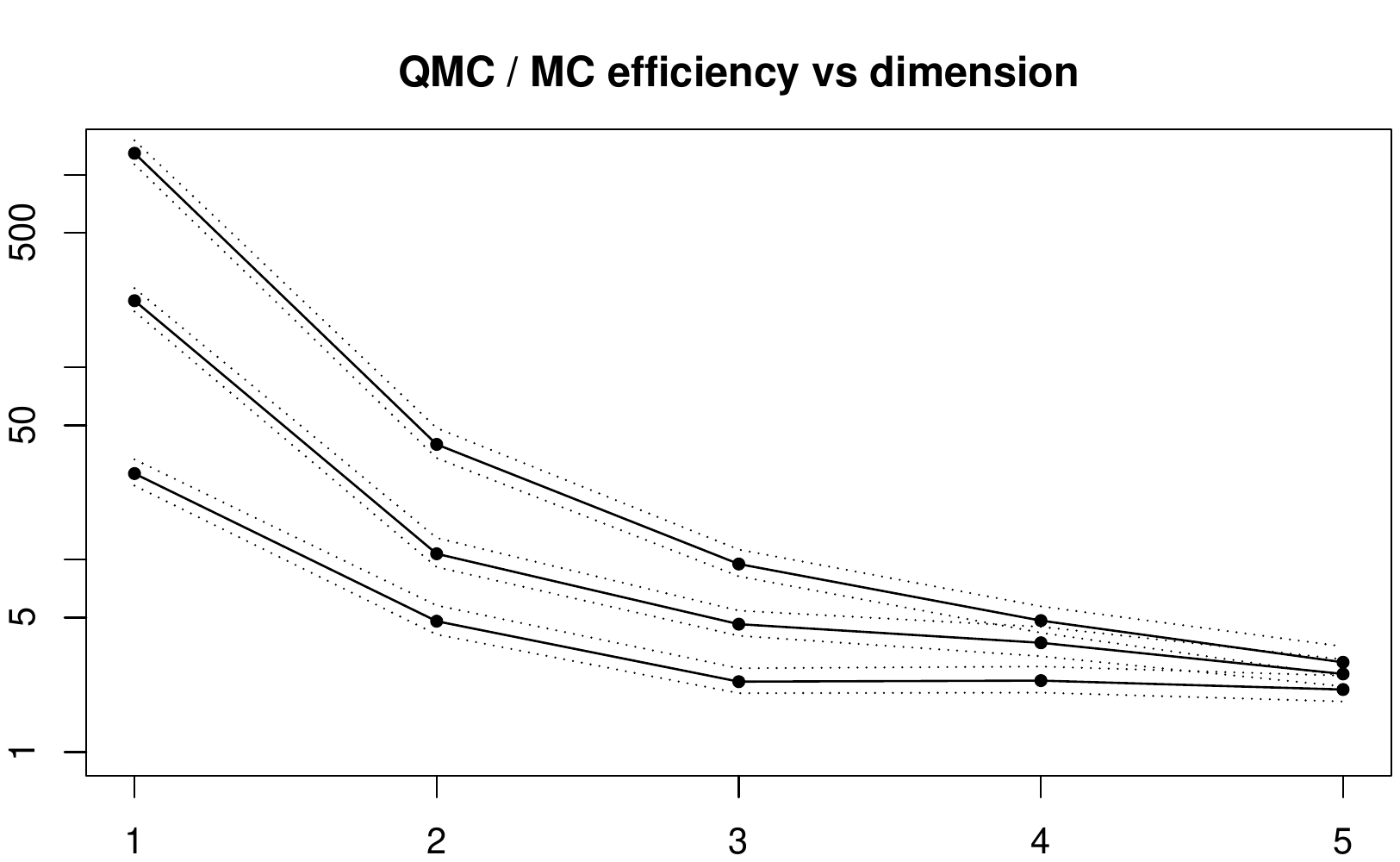}
\caption{\label{fig:case4}
RQMC efficiency for function $f_4$ based on $R=300$ repetitions.
Top to bottom $n=10^6$, $10^5$, $10^4$.
}
\end{figure}

The number $R$ of replications and the sample sizes
used varied with the difficulty of the case. 
Figures~\ref{fig:case1}, \ref{fig:case2} and \ref{fig:case3} show RQMC efficiency
with uncertainty bands for $f_1$, $f_2$ and $f_3$.
They are based on $R=100$ repetitions and $n$ from $100$ to $10{,}000$
by factors of $10$.
Figure~\ref{fig:case4} shows shows the rare event integrand $f_4$
with $R=300$ replications and $n$ from $10{,}000$ to $1{,}000{,}000$
by factors of $10$.

In Figure~\ref{fig:case1} we see dramatic efficiency gains versus
Monte Carlo.  As the sample size grows by $10$-fold the efficiency
grows also by about $10$-fold, when $d$ is small, consistent with a theoretically
anticipated MSE of $O(n^{-2+\epsilon})$.  The efficiency
decreases with dimension and by $d=50$ the efficiency no longer looks
to be proportional to $n$, but it is still quite large, e.g., over $100$
for the largest sample size.

Figure~\ref{fig:case2} has a discontinuous integrand.  Already by $d=10$
the efficencies are not much greater than $1$. The bands around the estimates
begin to overlap considerably as they must if all of the efficiencies are getting
closer to $1$.

Although the integrand $f_3$ has infinite variation for $d\ge3$,
the efficiencies in Figure~\ref{fig:case3} are stilll very large and grow
with sample size.
The smoothing effect from \cite{grie:kuo:sloa:2010} appears to apply
much more strongly to $f_3$ than to $f_2$.

The integrand $f_4$ is both discontinuous and describes a rare event
with $10^{-3}$ probability. Larger sample sizes and more repetitions and
fewer dimensions were used.  We see an efficiency gain but it diminishes
rapidly with dimension.

\subsection{Mean dimension}

Here we use the functional ANOVA of \cite{hoef:1948} and \cite{sobo:1967:tran}.
Let the function $f$ have functional ANOVA decomposition
$f(\bsx) = \sum_{u\subseteq1{:}d}f_u(\bsx)$
where $f_u$ depends on $\bsx$ only through
$x_j$ for $j\in u$. The effects $f_u$ have variance components
$\sigma^2_u = \var(f_u(\bsx))$.
The function $f$ is said to be of effective dimension $s$
in the superposition sense \citep{cafl:moro:owen:1997}, if
$$
\sum_{u:|u|\le s}\sigma^2_u \ge 0.99\sigma^2,
$$
and the same is not also true for $|u|\le s-1$.
This means that $99$\% or more of the variance comes from 
main effects $f_{\{j\}}$ and interactions of order up to $s$.
It is difficult to estimate the effective dimension of a function and a
more easily quantified measure is the mean dimension
\begin{align}\label{eq:meandim}
\bar d(f) = \frac{\sum_{u\subseteq1{:}d}|u|\sigma_u^2}{\sigma^2}.
\end{align}
See \cite{meandim}.

For any dimension $d\ge1$,  the test functions satisfy
$$
\mu = \int_{[0,1]^d}f(\bsx)\rd\bsx 
= \int_{-\infty}^\infty g(z)\varphi(z)\rd z 
$$
where $\varphi$ is the $\dnorm(0,1)$ probability density 
function. Similarly, for any $d\ge1$, the variance of the integrand is
$$
\sigma^2 = 
\int_{-\infty}^\infty (g(z)-\mu)^2\varphi(z)\rd z. 
$$

For $\bsx\in[0,1]^d$ and $z_j\in[0,1]$,
let $\bsx_{-\{j\}}{:}z_j$ be the point $\bsx$ after
$x_j$ has been  changed to $z_j$.
The theory of Sobol' indices \citep{sobo:1993} 
can be used to simplify the
numerator in the mean dimension~\eqref{eq:meandim},
\begin{align*}
\sum_{u\subseteq1{:}d}|u|\sigma^2_u 
&=\frac12\sum_{j=1}^d\int_0^1\int_{[0,1]^d} (f(\bsx)-f(\bsx_{-\{j\}}{:} z_j))^2\rd\bsx\rd z_j\\
&=\frac{d}2 
\int_0^1\int_{[0,1]^d} (f(\bsx)-f(x_{-\{1\}}{:}z_{1}))^2\rd\bsx\rd z_1\\
&=\frac{d}2\times 
\e\bigl([g(y_0+y_1)-g(y_0+y_2)]^2\bigr) 
\end{align*}
for independent $y_1,y_2\sim\dnorm(0,1/d)$
and $y_0\sim\dnorm(0,(d-1)/d)$. 
We may compute these integrals using
a randomized Halton point set in $3$ dimensions
for any given $d$.

\begin{figure}
\centering 
\includegraphics[width=.9\hsize]{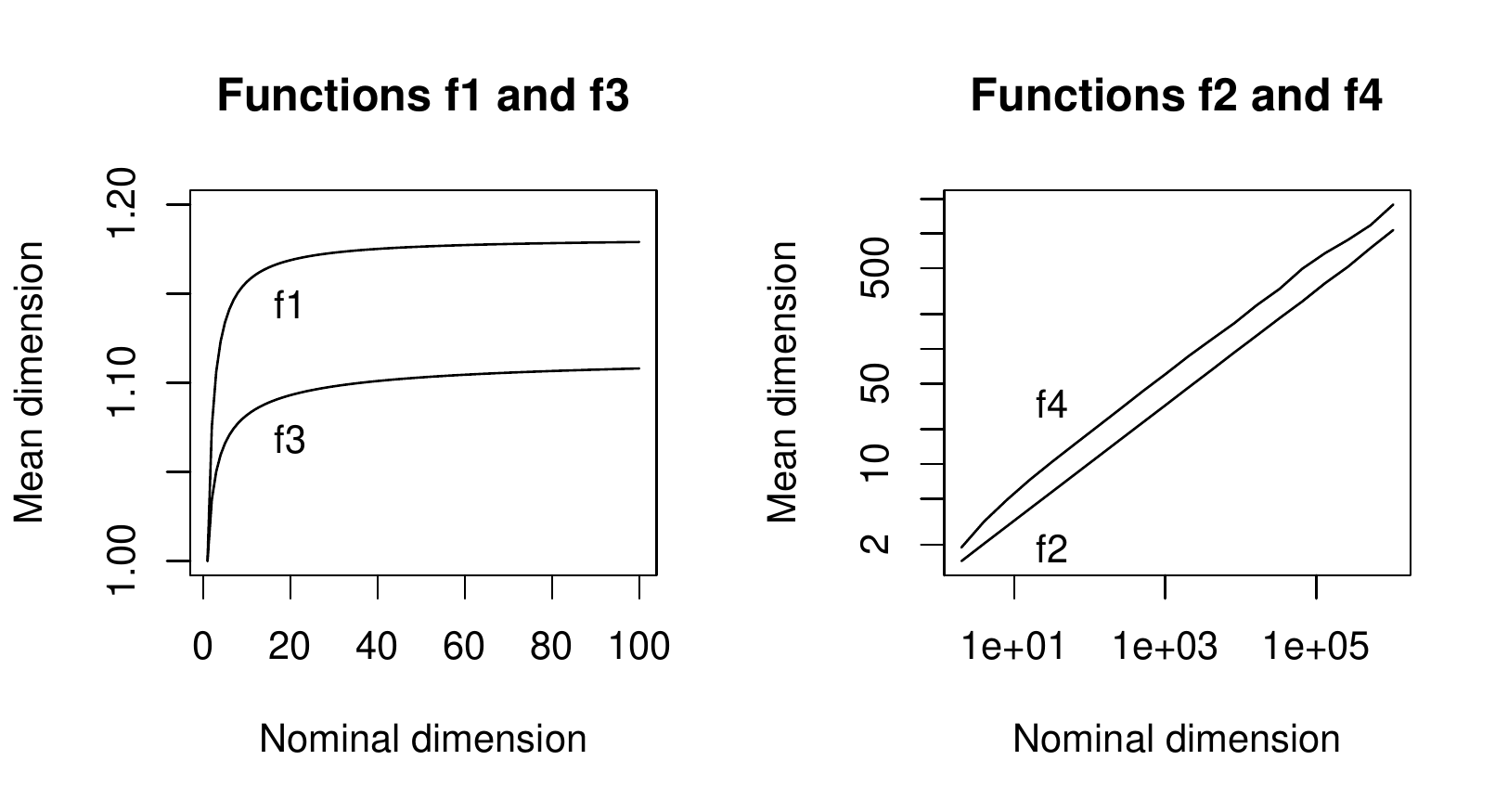}
\caption{\label{fig:meandimsa}
The left panel shows mean dimensions of $f_1$ and $f_3$
for $d=1,2,\dots,100$. The right panel shows mean dimensions 
of $f_2$ and $f_4$ for mean dimensions $2^m$ for $m=1,\dots,20$. 
}
\end{figure}

Figure~\ref{fig:meandimsa} shows the mean dimensions
for the test functions in this paper as a function of
the nominal dimension $d$.
The functions $f_1$ and $f_3$ have a mean dimension
that remains just barely larger than $1.0$ as $d\to\infty$.
The functions $f_2$ and $f_4$ have a mean dimension
that grows roughly like $\sqrt{d}$.

For very large $d$ and differentiable $g$ we anticipate that 
\begin{align*}
\sum_{u\subseteq1{:}d}|u|\sigma^2_u 
&\doteq \frac{d}2 
\times 
\e\bigl( g'(y_0)^2(y_1-y_2)^2\bigr) 
= 
\e\bigl( g'(y_0)^2\bigr),
\end{align*}
by independence of $y_0$, $y_1$ and $y_2$. 
So the mean dimension is approximately
$$
\frac{\int_{-\infty}^\infty g'(z)^2\varphi(z)\rd z}
{\int_{-\infty}^\infty (g(z)-\mu)^2\varphi(z)\rd z}
$$
for large $d$.  These limiting quantities closely match the
mean dimensions of $f_1$ and $f_3$ for very large values of $d$ such as $10^6$.

\section{Comparisons}\label{sec:comparison}

There is a randomized Halton code function {\tt ghalton}
in the \proglang{R} package {\tt qrng} by \cite{hofe:lemi:2016}.
That code uses a random linear scramble
and it has core computations written in \proglang{C++}.

\begin{figure}
\centering 
\includegraphics[width=.9\hsize]{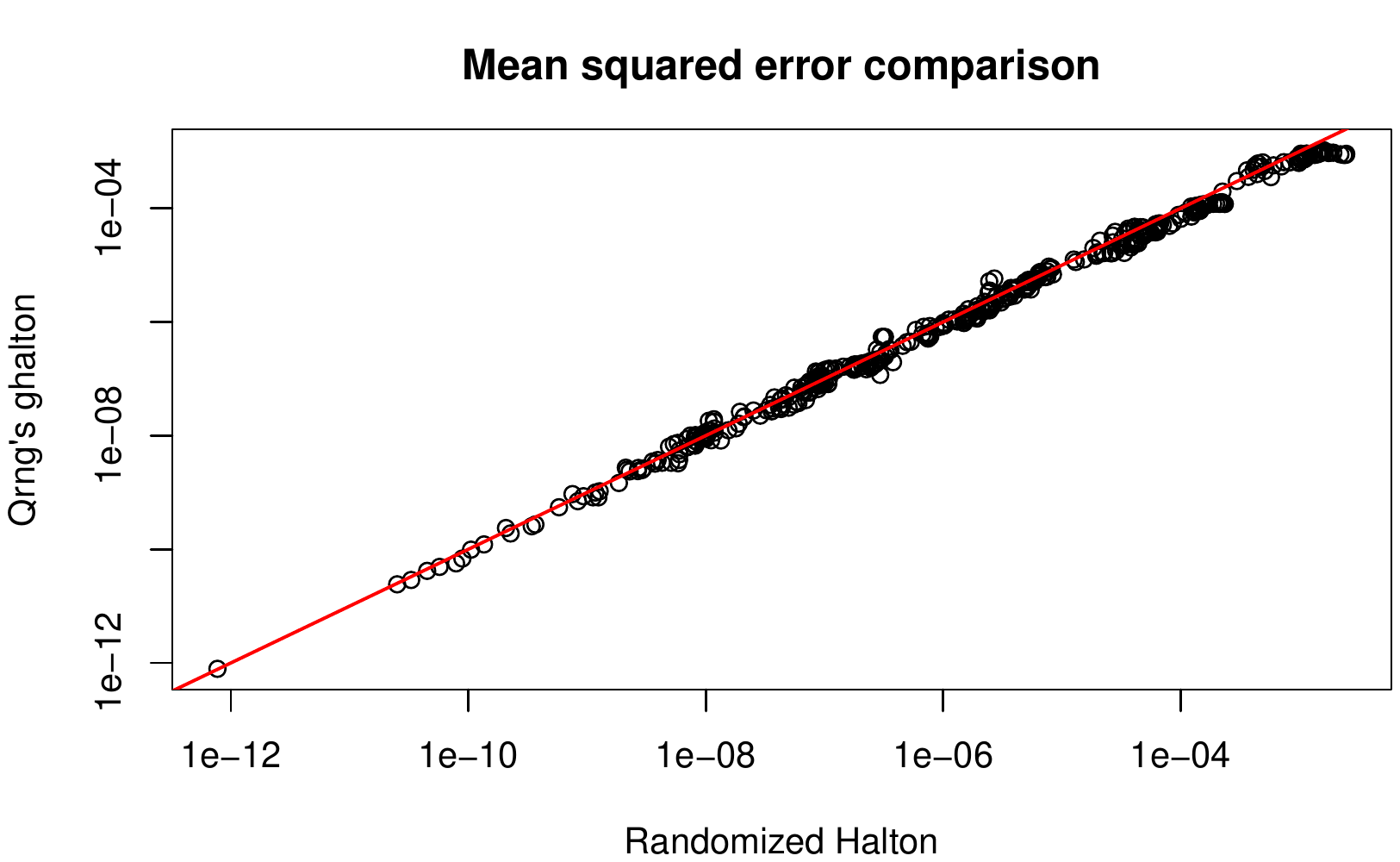}
\caption{\label{fig:rhrvsqrng}
The horizontal axis has mean squared errors from {\tt rhalton}.
The vertical axis has mean squared errors for the {\tt ghalton}
program from the \proglang{R} package \pkg{qrng}. There is
a reference line where the MSEs are equal.
}
\end{figure}

The problems displayed in Figures~\ref{fig:case1} through~\ref{fig:case4}
were also computed with the {\tt ghalton} generalized Halton code from
the {\tt qrng} package.
Figure~\ref{fig:rhrvsqrng} depicts the mean squared errors from
each method on all of the dimensions and sample sizes in those
figures.  There is not much difference in accuracy.
That figure includes data with very large differences in sample
sizes as well as differences in efficiencies, and the mean squared
errors in each method range by a factor of more than $10^8$.
We might also want to compare efficiencies, normalizing out
the large differences among sample sizes.  Sample size effects cancel in the average
of MSE({\tt rhalton})/MSE({\tt ghalton}) which is $1.22$
and in the average of the reciprocal
 MSE({\tt ghalton})/MSE({\tt rhalton}) is $0.88$.
Both of these reflect an efficiency advantage for {\tt ghalton}, but not a large one.

The {\tt ghalton} code does not come with internal 
seeding methods. 
Version 0.0-3 allows up to $d=360$ dimensions.
If the seed is set and $n$ and $d$ are increased,
the new matrix contains the previous one as its
upper left corner:
\begin{verbatim}
> set.seed(1);x1=ghalton(500,180);set.seed(1);x2=ghalton(1000,360)
> range(x2[1:500,1:180]-x1)
[1] 0 0
\end{verbatim}

The principal advantage of the present {\tt rhalton} code is that it
can be extended to $n'>n$ rows only computing the new $n'-n$
rows, or to $d'>d$ columns only computing the new $d'-d$ columns.
It is even possible to extract an arbitrary single element  $\cx^{(s)}_{ij}$
via {\tt rhalton(1,1,n0=i-1,d0=j-1,singleseed=s)}.
A second advantage is that the pseudocode is so simple that
one can easily write it in a new language, such as \proglang{julia}
\citep{beza:edel:karp:shah:2017},
or use it for instructional purposes about RQMC.

There are more accurate RQMC methods than this one,
if the integrand has sufficient smoothness.  
See, for instance \cite{dick:2011} or \cite{smoovar}.
However any
rule with error or root mean squared error $o(1/n)$ cannot
be reasonably used at arbitrary sample sizes.
To see why, suppose that $\hat\mu_n$ is an average of $n$
evaluations and $\hat \mu_{n+1}$ averages $n+1$ of them.
Then $\hat\mu_{n+1}-\hat\mu_n=O(1/n)$. If $\hat\mu_n$
had error $o(1/n)$ then we have probably made it an order of magnitude
worse just by taking one more observation. The only exception would
be if $f(x_{n+1})$ itself had error $o(1/n)$.  In that case we would
just use $f(x_{n+1})$ all by itself. This observation is due to \cite{sobo:1998} and 
was worked out in detail in \cite{quadconstraint}. The consequence is
that for $o(1/n)$ accuracy an equal weight rule must use sample
sizes $n_\ell$ with some lower bound on $n_{\ell+1}/n_\ell$.
Equal weight rules with arbitrary sample sizes are easiest to use.

Nested uniform scrambling of $(t,m,d)$-nets in base $b$ \citep{rtms}
has the property that for finite $n$ the resulting variance
cannot be much worse than plain MC for any integrand $f$ in $L^2$.
The bound depends on $t,m,d$ and $b$ for sample sizes $n=b^m$.
The randomized Halton method presented
here does not have that property.  Pathological exceptions are possible.
Suppose that $d=26$ and that $f(x_i)$ only depends on the $26$'th
component $x_{ij}$. The $26$'th prime is $101$ and to make matters
worse, suppose that $f(x_i)$ is a periodic function of $x_{i,26}$
with period $1/101$.  Then the first $101$ values of $f(x_i)$ are all equal.
So are the second $101$ values. The variance could be $101$ times
as large as Monte Carlo sampling, and of course a larger prime
number would be even more problematic, as would a function
that depended only on $x_j$ and had period $p_j^k$ for some $k$.
The code presented here is based on ignoring this far-fetched
possibility for the sake of simplicity of implementation and use.
\section*{Acknowledgments}

This work was supported by the NSF under
grants DMS-1407397, and DMS-1521145.
\bibliographystyle{apalike}
\bibliography{qmc}
\vfill\eject 
\section*{Appendix: source code in R}
\begin{verbatim}
#
# Coded by Art B. Owen, Stanford University.
# April 2017
#
rhalton <- function(n,d,n0=0,d0=0,singleseed,seedvector){
#
# Randomly scrambled Halton sequence of n points in d dimensions.
#
# If you already have n0 old points, set n0 to get the next n points.
# If you already have d0 old components, set d0 to get the next d inputs.
# Get points  n0 + 0:(n-1)  in dimensions  d0 + (1:d)
#
# seedvector  = optional vector of  d0+d random seeds, one per dimension
# singleseed  = optional scalar seed (has lower precedence than seedvector)

# Handle input dimension correctness and corner cases
  if( min(n0,d0) < 0 )
    stop(paste("Starting indices (n0, d0) cannot be < 0, input had (",n0,",",d0,")"))
  if( min(n,d) < 0 )
    stop(paste("Cannot have negative n or d"))
  if( n==0 || d==0 ) # Odd corner cases: user wants n x 0 or 0 x d matrix.
    return(matrix(nrow=n,ncol=d))
  D <- nthprime(0,getlength=TRUE)
  if( d0+d > D )
    stop(paste("Implemented only for d <=",D))

# Seed rules
  if( !missing(singleseed) && length(singleseed) != 1 )
    stop("singleseed, if supplied, must be scalar")
  if( !missing(seedvector) && length(seedvector) < d0+d )
    stop("seedvector, if supplied, must be have length at least d0+d")
  if(  missing(seedvector) && !missing(singleseed) )
    seedvector <- singleseed + 1:(d0+d) - 1

# Generate and return the points
  ans <- matrix(0,n,d)
  for( j in 1:d ){  
    dimj <- d0+j
    if( !missing(seedvector) )set.seed(seedvector[dimj])
    ans[,j] <- randradinv( n0 + 0:(n-1), nthprime(dimj) )  # zero indexed rows
  }
  ans
}

randradinv <- function(ind,b=2){
# Randomized radical inverse functions for indices in ind and for base b.
# The calling routine should set the random seed if reproducibility is desired.

  if(any(ind!=round(ind)))
    stop("Indices have to be integral.")
  if(any(ind<0))
    stop("Indices cannot be negative.")
  if(!(is.vector(b) && length(b)==1 && b==floor(b) && b>1))
    stop("b must be a single integer >= 2")

  b2r <- 1/b
  ans <- ind*0
  res <- ind
  while( 1-b2r < 1 ){  # Assumes floating point comparisons, fixed precision.
    dig  <- res %% b
    perm <- sample(b)-1
    pdig <- perm[1+dig]
    ans  <- ans + pdig * b2r
    b2r  <- b2r/b
    res  <- (res - dig)/b
  }
  ans
}

nthprime <- function(n,getlength=FALSE){
# Get the n'th prime, using a hard coded list (or get the length of that list).

# First fifty-eight prime numbers from OEIS A000040, April 2017
primes <- c( 2,   3,   5,   7,  11,  13,  17,  19,  23,  29, 
            31,  37,  41,  43,  47,  53,  59,  61,  67,  71, 
            73,  79,  83,  89,  97, 101, 103, 107, 109, 113, 
           127, 131, 137, 139, 149, 151, 157, 163, 167, 173, 
           179, 181, 191, 193, 197, 199, 211, 223, 227, 229, 
           233, 239, 241, 251, 257, 263, 269, 271)

# First 1000 prime numbers from primes.utm.edu/lists/small/1000.txt, April 2017
primes <- c(
     2,    3,    5,    7,   11,   13,   17,   19,   23,   29 
,   31,   37,   41,   43,   47,   53,   59,   61,   67,   71 
,   73,   79,   83,   89,   97,  101,  103,  107,  109,  113 
,  127,  131,  137,  139,  149,  151,  157,  163,  167,  173 
,  179,  181,  191,  193,  197,  199,  211,  223,  227,  229 
,  233,  239,  241,  251,  257,  263,  269,  271,  277,  281 
,  283,  293,  307,  311,  313,  317,  331,  337,  347,  349 
,  353,  359,  367,  373,  379,  383,  389,  397,  401,  409 
,  419,  421,  431,  433,  439,  443,  449,  457,  461,  463 
,  467,  479,  487,  491,  499,  503,  509,  521,  523,  541 
,  547,  557,  563,  569,  571,  577,  587,  593,  599,  601 
,  607,  613,  617,  619,  631,  641,  643,  647,  653,  659 
,  661,  673,  677,  683,  691,  701,  709,  719,  727,  733 
,  739,  743,  751,  757,  761,  769,  773,  787,  797,  809 
,  811,  821,  823,  827,  829,  839,  853,  857,  859,  863 
,  877,  881,  883,  887,  907,  911,  919,  929,  937,  941 
,  947,  953,  967,  971,  977,  983,  991,  997, 1009, 1013 
, 1019, 1021, 1031, 1033, 1039, 1049, 1051, 1061, 1063, 1069 
, 1087, 1091, 1093, 1097, 1103, 1109, 1117, 1123, 1129, 1151 
, 1153, 1163, 1171, 1181, 1187, 1193, 1201, 1213, 1217, 1223 
, 1229, 1231, 1237, 1249, 1259, 1277, 1279, 1283, 1289, 1291 
, 1297, 1301, 1303, 1307, 1319, 1321, 1327, 1361, 1367, 1373 
, 1381, 1399, 1409, 1423, 1427, 1429, 1433, 1439, 1447, 1451 
, 1453, 1459, 1471, 1481, 1483, 1487, 1489, 1493, 1499, 1511 
, 1523, 1531, 1543, 1549, 1553, 1559, 1567, 1571, 1579, 1583 
, 1597, 1601, 1607, 1609, 1613, 1619, 1621, 1627, 1637, 1657 
, 1663, 1667, 1669, 1693, 1697, 1699, 1709, 1721, 1723, 1733 
, 1741, 1747, 1753, 1759, 1777, 1783, 1787, 1789, 1801, 1811 
, 1823, 1831, 1847, 1861, 1867, 1871, 1873, 1877, 1879, 1889 
, 1901, 1907, 1913, 1931, 1933, 1949, 1951, 1973, 1979, 1987 
, 1993, 1997, 1999, 2003, 2011, 2017, 2027, 2029, 2039, 2053 
, 2063, 2069, 2081, 2083, 2087, 2089, 2099, 2111, 2113, 2129 
, 2131, 2137, 2141, 2143, 2153, 2161, 2179, 2203, 2207, 2213 
, 2221, 2237, 2239, 2243, 2251, 2267, 2269, 2273, 2281, 2287 
, 2293, 2297, 2309, 2311, 2333, 2339, 2341, 2347, 2351, 2357 
, 2371, 2377, 2381, 2383, 2389, 2393, 2399, 2411, 2417, 2423 
, 2437, 2441, 2447, 2459, 2467, 2473, 2477, 2503, 2521, 2531 
, 2539, 2543, 2549, 2551, 2557, 2579, 2591, 2593, 2609, 2617 
, 2621, 2633, 2647, 2657, 2659, 2663, 2671, 2677, 2683, 2687 
, 2689, 2693, 2699, 2707, 2711, 2713, 2719, 2729, 2731, 2741 
, 2749, 2753, 2767, 2777, 2789, 2791, 2797, 2801, 2803, 2819 
, 2833, 2837, 2843, 2851, 2857, 2861, 2879, 2887, 2897, 2903 
, 2909, 2917, 2927, 2939, 2953, 2957, 2963, 2969, 2971, 2999 
, 3001, 3011, 3019, 3023, 3037, 3041, 3049, 3061, 3067, 3079 
, 3083, 3089, 3109, 3119, 3121, 3137, 3163, 3167, 3169, 3181 
, 3187, 3191, 3203, 3209, 3217, 3221, 3229, 3251, 3253, 3257 
, 3259, 3271, 3299, 3301, 3307, 3313, 3319, 3323, 3329, 3331 
, 3343, 3347, 3359, 3361, 3371, 3373, 3389, 3391, 3407, 3413 
, 3433, 3449, 3457, 3461, 3463, 3467, 3469, 3491, 3499, 3511 
, 3517, 3527, 3529, 3533, 3539, 3541, 3547, 3557, 3559, 3571 
, 3581, 3583, 3593, 3607, 3613, 3617, 3623, 3631, 3637, 3643 
, 3659, 3671, 3673, 3677, 3691, 3697, 3701, 3709, 3719, 3727 
, 3733, 3739, 3761, 3767, 3769, 3779, 3793, 3797, 3803, 3821 
, 3823, 3833, 3847, 3851, 3853, 3863, 3877, 3881, 3889, 3907 
, 3911, 3917, 3919, 3923, 3929, 3931, 3943, 3947, 3967, 3989 
, 4001, 4003, 4007, 4013, 4019, 4021, 4027, 4049, 4051, 4057 
, 4073, 4079, 4091, 4093, 4099, 4111, 4127, 4129, 4133, 4139 
, 4153, 4157, 4159, 4177, 4201, 4211, 4217, 4219, 4229, 4231 
, 4241, 4243, 4253, 4259, 4261, 4271, 4273, 4283, 4289, 4297 
, 4327, 4337, 4339, 4349, 4357, 4363, 4373, 4391, 4397, 4409 
, 4421, 4423, 4441, 4447, 4451, 4457, 4463, 4481, 4483, 4493 
, 4507, 4513, 4517, 4519, 4523, 4547, 4549, 4561, 4567, 4583 
, 4591, 4597, 4603, 4621, 4637, 4639, 4643, 4649, 4651, 4657 
, 4663, 4673, 4679, 4691, 4703, 4721, 4723, 4729, 4733, 4751 
, 4759, 4783, 4787, 4789, 4793, 4799, 4801, 4813, 4817, 4831 
, 4861, 4871, 4877, 4889, 4903, 4909, 4919, 4931, 4933, 4937 
, 4943, 4951, 4957, 4967, 4969, 4973, 4987, 4993, 4999, 5003 
, 5009, 5011, 5021, 5023, 5039, 5051, 5059, 5077, 5081, 5087 
, 5099, 5101, 5107, 5113, 5119, 5147, 5153, 5167, 5171, 5179 
, 5189, 5197, 5209, 5227, 5231, 5233, 5237, 5261, 5273, 5279 
, 5281, 5297, 5303, 5309, 5323, 5333, 5347, 5351, 5381, 5387 
, 5393, 5399, 5407, 5413, 5417, 5419, 5431, 5437, 5441, 5443 
, 5449, 5471, 5477, 5479, 5483, 5501, 5503, 5507, 5519, 5521 
, 5527, 5531, 5557, 5563, 5569, 5573, 5581, 5591, 5623, 5639 
, 5641, 5647, 5651, 5653, 5657, 5659, 5669, 5683, 5689, 5693 
, 5701, 5711, 5717, 5737, 5741, 5743, 5749, 5779, 5783, 5791 
, 5801, 5807, 5813, 5821, 5827, 5839, 5843, 5849, 5851, 5857 
, 5861, 5867, 5869, 5879, 5881, 5897, 5903, 5923, 5927, 5939 
, 5953, 5981, 5987, 6007, 6011, 6029, 6037, 6043, 6047, 6053 
, 6067, 6073, 6079, 6089, 6091, 6101, 6113, 6121, 6131, 6133 
, 6143, 6151, 6163, 6173, 6197, 6199, 6203, 6211, 6217, 6221 
, 6229, 6247, 6257, 6263, 6269, 6271, 6277, 6287, 6299, 6301 
, 6311, 6317, 6323, 6329, 6337, 6343, 6353, 6359, 6361, 6367 
, 6373, 6379, 6389, 6397, 6421, 6427, 6449, 6451, 6469, 6473 
, 6481, 6491, 6521, 6529, 6547, 6551, 6553, 6563, 6569, 6571 
, 6577, 6581, 6599, 6607, 6619, 6637, 6653, 6659, 6661, 6673 
, 6679, 6689, 6691, 6701, 6703, 6709, 6719, 6733, 6737, 6761 
, 6763, 6779, 6781, 6791, 6793, 6803, 6823, 6827, 6829, 6833 
, 6841, 6857, 6863, 6869, 6871, 6883, 6899, 6907, 6911, 6917 
, 6947, 6949, 6959, 6961, 6967, 6971, 6977, 6983, 6991, 6997 
, 7001, 7013, 7019, 7027, 7039, 7043, 7057, 7069, 7079, 7103 
, 7109, 7121, 7127, 7129, 7151, 7159, 7177, 7187, 7193, 7207
, 7211, 7213, 7219, 7229, 7237, 7243, 7247, 7253, 7283, 7297 
, 7307, 7309, 7321, 7331, 7333, 7349, 7351, 7369, 7393, 7411 
, 7417, 7433, 7451, 7457, 7459, 7477, 7481, 7487, 7489, 7499 
, 7507, 7517, 7523, 7529, 7537, 7541, 7547, 7549, 7559, 7561 
, 7573, 7577, 7583, 7589, 7591, 7603, 7607, 7621, 7639, 7643 
, 7649, 7669, 7673, 7681, 7687, 7691, 7699, 7703, 7717, 7723 
, 7727, 7741, 7753, 7757, 7759, 7789, 7793, 7817, 7823, 7829 
, 7841, 7853, 7867, 7873, 7877, 7879, 7883, 7901, 7907, 7919
)

p <- length(primes)
if( getlength )
  return(p)
if( any(n<1) )
  stop(paste(sep="","The n'th prime is only available for n >= 1, not n = ",min(n)))
if( any(n>p) )
  stop(paste(sep="","The present list of primes has length ",p,
   ". It does not include the n'th one for n = ",max(n),"."))

return(primes[n])
}




\end{verbatim}
\end{document}